\PassOptionsToPackage{table}{xcolor}
 \documentclass[useAMS,usenatbib,twocolumn]{aastex631}

 \usepackage{amsmath}
 \usepackage{amssymb}
 \usepackage{natbib}
 \usepackage{multirow}
 \usepackage{graphicx}\graphicspath{{graphics/}}
 \usepackage{color}
 \usepackage{epstopdf}

\received{\today}
\revised{\today}
\accepted{\today}
\submitjournal{ApJ}

\shorttitle{Galaxy Quenching at high redshifts}
\shortauthors{Remus \& Kimmig}

\begin{document}

\title{Relight the Candle: What happens to High Redshift Massive Quenched Galaxies}

\correspondingauthor{Rhea-Silvia Remus}
\email{rhea@usm.lmu.de}
\author{Rhea-Silvia Remus}
\affil{Universit\"ats-Sternwarte M\"unchen, Fakult\"at f\"ur Physik, Ludwig-Maximilians Universit\"at, Scheinerstr.\ 1, D-81679 M\"unchen, Germany}
\author{Lucas C. Kimmig}
\affil{Universit\"ats-Sternwarte M\"unchen, Fakult\"at f\"ur Physik, Ludwig-Maximilians Universit\"at, Scheinerstr.\ 1, D-81679 M\"unchen, Germany}

\begin{abstract}
A puzzling population of extremely massive quiescent galaxies at redshifts beyond z=3 has recently been revealed by JWST and ALMA, some of them with stellar ages that show their quenching times to be as high as z=6, while their stellar masses are already above $5\times10^{10}M_\odot$. These extremely massive yet quenched galaxies challenge our understanding of galaxy formation at the earliest stages. Using the hydrodynamical cosmological simulation suite Magneticum Pathfinder, we show that such massive quenched galaxies at high redshifts can be successfully reproduced with similar number densities as observed. The stellar masses, sizes, formation redshifts, and star formation histories of the simulated quenched galaxies match those determined with JWST. Following these quenched galaxies at $z=3.4$ forward in time, we find 20\% to be accreted onto a more massive structure by $z=2$, and from the remaining 80\% about 30\% rejuvenate up to $z=2$, another 30\% stay quenched, and the remaining 40\% rejuvenated on a very low level of star formation. Stars formed through rejuvenation are mostly formed on the outer regions of the galaxies, not in the centres. Furthermore, we demonstrate that the massive quenched galaxies do not reside in the most massive nodes of the cosmic web, but rather live in side-nodes of approximately Milky-Way halo mass. Even at $z=0$, only about 10\% end up in small-mass galaxy clusters, while most of the quenched galaxies at $z=3.4$ end up in group-mass halos, with about 20\% actually not even reaching $10^{13}M_\odot$ in halo mass.
\end{abstract}


\section{Introduction}
The high-redshift Universe has become available to observations in the last year thanks to JWST in unprecedented detail, revealing star forming galaxies to exist at redshifts as high as $z=13$ or possibly higher from photometry \citep[e.g.][]{adams:2023,harikane:2023}, and despite some of them having shown to be at lower redshifts by spectroscopic follow ups, several of them have been confirmed at redshifts as high as $z=13$ \citep[e.g.][]{harikane:2023spec,arrabal:2023}. Furthermore, ALMA has added to the zoo of high redshift galaxies observations of low-turbulence but dust-obscured disk galaxies that cannot be detected by JWST due to the high dust content \citep[e.g.][]{lelli:2023,rizzo:2023}. Such dust obscured galaxies with average star formation rates have been found even up to $z=7$ \citep{fudamoto:2021}. Massive galaxies with disk morphologies have been resolved at redshifts up to at least $z=4$ \citep[e.g.,][]{tsukui:2021,roman:2023}, with star formation rates of more than $1000M_*/yr$ and strong AGN activity \citep{tsukui:2023}, but also some spiral galaxies have been reported at those redshifts that show signs of their star formation being shut off \citep{fudamoto:2022,nelson:2023}. All of these observations challenge our understanding of galaxy formation and the overall structure formation and cosmology when the Universe was very young, and are the perfect testbed for theoretical models and simulations to probe our grasp on the underlying physics.

To this zoo of extraordinary galaxies at high redshift also belong massive quenched galaxies, that is galaxies at redshifts as high as $z=4$ with stellar masses of $M_*>3\times10^{11}M_\odot$ \citep{schreiber:2018,nanayakkara:2022,carnall:2023,long:2023}, for which spectroscopically their formation times, i.e. the time at which half of their stars have been formed, were confirmed to be as high as $z=6-7$ \citep{nanayakkara:2022,carnall:2023}, in one case even as high as $z=11$ \citep{glazebrook:2023}.
These quiescent galaxies are found in significant numbers at high redshift, unexpectedly, as with the overall gas content in these galaxies much higher than at present day, it is still a matter of debate how these galaxies have come to be quenched at such early times, but also how they evolve after they were quenched, and what they become at $z=0$.

Quenched, or quiescent, galaxies at high redshifts are known to be more compact than their present-day counterparts of the same stellar mass, more so the higher the redshift \citep[e.g.][]{vdwel:2014,ito:2023}. Such compact quiescent galaxies are not observed at low redshifts, at least not in quantities that can account for the numbers of compact quenched galaxies now reported at high redshifts. Multiple dry minor mergers have been proposed as a possible formation pathway for such quenched galaxies \citep{bournaud:2007,bezanson:2009,naab:2009}, as such small mergers deposit most of their stellar mass at larger radii and as such enhance the radial growth compared to normal major merger events \citep{hilz:2012,karademir:2019}. However, dry mergers are known to be more common the lower the redshift \citep[e.g.][]{bell:2006}, and rather rare at high redshifts before $z=2$, where the gas fractions are generally much higher. 
Thus, other additional evolution pathways for such high redshift quenched galaxies are required. One possible evolution pathway is for the compact quenched galaxy to merge into another, more massive (and possibly still star forming) galaxy as the minor progenitor. 

Another suggested pathway is so-called rejuvination, that is re-accretion of gas from either the cosmic web or through gas-rich mergers onto the central galaxy, thereby reforming a disk and re-starting star formation, Indications for such a process have been found imprinted in observed quiescent galaxies at present-day \citep[e.g.][]{yi:2005}, clearly showing that the simple assumption that once a galaxy has been quenched it stays quenched is not valid. Similarly, about 16\% of the quiescent galaxies found at $z\approx0.8$ in the LEGA-C survey show signs of rejuvination \citep{chauke:2019}. Models predict a range of 10\% to 70\% of all massive galaxies to have lived through a rejuvination phase \citep{zhang:2023}, and even the possibility of multiple rejuvination events for a galaxy throughout its life \citep{tanaka:2023}. The exact conditions under which some quenched galaxies can revive through accretion of new gas and others do not are yet not well understood. However, observations of local galaxies that host AGN show signs of their last massive star burst event to have occured at the same time as the last AGN activity, indicating that both AGN activity and the massive starburst might have been triggered by the same event \citep{martinnavarro:2022}.

Using the Illustris-TNG300 simulation, \citet{hartley:2023} identified 5 quenched galaxies emerging at $z=4.2$, which were all quenched by the implemented AGN feedback and ended up as the most massive nodes of the box at $z=0$. However, their predicted number of quenched galaxies is lower than what has been observed by \citet{long:2023,carnall:2023}. More problematic, observations also report some of the quenched galaxies to have been quenched since $z=5$ or earlier \citep{nanayakkara:2022,carnall:2023}, which is higher than found for any quenched galaxy in that simulation \citep{kakimoto:2023}. Thus, many questions are actually still open, which we will address in this study together with its companion study by Kimmig et al., submitted, hereafter K23.

This paper will focus on the properties and future evolution pathways of quenched galaxies found in the Magneticum simulations, and compare the properties of those galaxies to observations. The companion paper K23 focuses on the mechanisms and environmental conditions that lead to the quenching, discussing how the feedback both from the stars and the central supermassive black hole combined are responsible. 
The paper is structures as follows: Sec.~\ref{sec:sim} will introduce the simulation suite used to study quenched galaxies at high redshifts. Sec~\ref{sec:quenched} presents the sample of quiescent galaxies, and compares global properties between quenches simulated and observed galaxies, including the formation redshifts obtained from stellar populations. Sec.~\ref{sec:future} focusses on the different pathways galaxies follow after they have been quenched at high redshifts. In particular, we discuss the process of rejuvination, with an outlook to the halo properties of the quenched galaxies at $z=0$. Finally, Sec.~\ref{sec:conclusion} will summarize the results and discuss them in the global context of galaxy evolution since cosmic dawn. 

\section{Simulations and Observations}\label{sec:sim}
Finding quenched galaxies at high redshifts requires a box volume large enough to encompass all potential environments while at the same time the simulation needs to have high enough resolution to properly resolve galaxies with stellar masses larger than $1\times10^{10}M_\odot$ to capture the observed mass ranges of quiescent galaxies. We use Box3 with a volume of $(128Mpc/h)^3$ from the fully hydrodynamical cosmological simulation suite Magneticum Pathfinder (www.magneticum.org), in the ultra-high resolution uhr, with dark matter, gas, and stellar particles masses of $m_\mathrm{dm}=3.6\times10^{7} M_{\odot}/h$, $m_\mathrm{gas} = 7.3\times10^{6} M_{\odot}/h$, and $m_\mathrm{*}\approx1/4^{\rm th}m_\mathrm{gas}\approx1.8\times10^{6} M_{\odot}/h$, respectively, as every gas particle spawns up to 4 stellar particles. Gravitational softening for the three components is included as $\epsilon_\mathrm{dm} = \epsilon_\mathrm{gas} = 1.4~\mathrm{kpc}/h$ and $\epsilon_\mathrm{*} = 0.7~\mathrm{kpc}/h$. 
The simulation uses a WMAP-7 cosmology following \citet{komatsu:2011}, with $h=0.704$, $\Omega_m = 0.272$, $\Omega_b = 0.0451$, $\Omega_\lambda = 0.728$, $\sigma_8 = 0.809$,
and an initial slope of the power spectrum of $n_s = 0.963$. 

The simulation was performed with an updated version of GADGET-2 \citep{springel:2005}, with modifications to SPH (including artificial viscosity and thermal conductivity) according to \citep{dolag:2004,dolag:2005,donnert:2013,beck:2015}. It includes full baryonic physics as described in detail by \citet{teklu:2015}. Star formation and stellar feedback are implemented based on the model by \citet{springel:2003}. Stellar feedback encompasses feedback from SN-I, SN-II, and AGB stars, with metal enrichment and cooling from all three sources are implemented according to \citet{tornatore:2004,tornatore:2007} and \citet{wiersma:2009}. Detailed descriptions of the metal enrichment model are given by \citet{dolag:2017}. Black hole treatment and feedback is included as described by \citet{steinborn:2015} based on the model by \citet{fabjan:2010}. In addition, UV/X-ray background and CMB background radiation are computed according to \citet{haardt:2001}.

Galaxies are identified using SUBFIND \citep{springel:2001, dolag:2009}. Galaxies that subfind identifies as the most massive galaxy in a given halo are called ``central'' galaxies, while we call all galaxies that are not the centrals of their host halo ``satellite'' galaxies.
Galaxies are traced forward and backwards in time using L-BaseTree \citep{springel:2005b}. Towards higher redshifts, the most massive galaxy at each snapshot is used as the main progenitor, and that branch of the merger tree is followed. Tracing forward, we mark those galaxies that merge with a larger galaxy, i.e. that are the smaller partner in the merger event, as ``central subs'', since the remnant galaxy after the merger is still a central, but the galaxy we were tracing was the minor contributor to the mass growth and thus not the main branch itself. This procedure works until the last high-resolution snapshot at $z\approx2$. To trace halos at redshifts below $z\approx2$, we match halos at the last available snapshot to the halos in the simulation Box3 hr, which ran until z=0 with the same initial conditions but a lower resolution, and track those forward to z=0. Due to the lower resolution, we only use global information like stellar or total mass, and merger mass ratios, to track down to z=0.

The simulation Box3 uhr ran until $z\approx 2$, covering the redshift evolution with outputs since $z\approx15$. Most importantly, it provides a snapshot with full particle information at $z=3.4$, which is comparable to the average redshift at which the quenched galaxies presented by \citet{nanayakkara:2022} or \citet{long:2023} are observed. At $z=3.4$, $2903$ galaxies are found with stellar masses above $M_*>2\times10^{10}M_\odot$, above which we consider galaxies to be resolved well enough to be included in this study for more than pure number counting aspects.

\section{Quenched Galaxies at $z\approx3.4$}\label{sec:quenched}
Observed quiescent galaxies with full spectra have recently been reported by \citet{nanayakkara:2022} and \citet{carnall:2023} using JWST NIRIS spectra at redshifts of $z\approx 3-4$, enabling to examine the stellar age distributions and thus the quenching and formation times of these galaxies. In the following, we identify quenched quiescent galaxies at a redshift of $z=3.4$ in the simulations and compare their global properties to their observed counterparts.

Throughout this work we consider galaxies with 3 different stellar mass cuts: I) a stellar mass of $M_*\geq 2\times 10^{10}M_\odot$, for which star formation properties are well enough resolved to safely identify quiescent galaxies; II) a stellar mass of  $M_*\geq 3\times 10^{10}M_\odot$, where these are well resolved with more than $10000$~stellar particles; III) a stellar mass of $M_*\geq 5\times 10^{10}M_\odot$, encompassing the best resolved and most massive galaxies at the given redshift. At $z=3.42$, this encompasses $2903$ galaxies in cut I), and $1309$~galaxies above the most important cut level II) of which $1217$~are centrals and $92$~are satellite galaxies. Above the third cut range, cut III), we find $395$ galaxies.
\begin{figure}
  \begin{center}
    \includegraphics[width=\columnwidth]{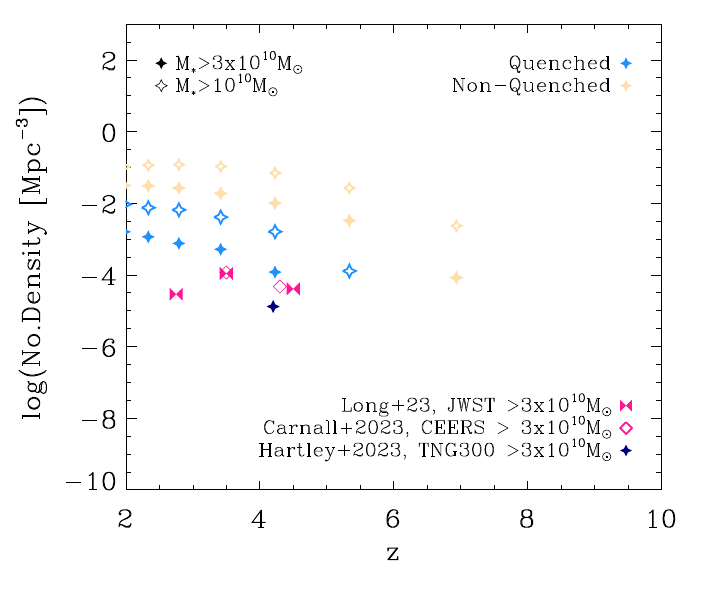}
    \caption{
    Number densities of quenched (blue) and non-quenched (sand) galaxies with stellar masses above $3\times10 ^{10}M_\odot$ (filled) and $1\times10^{10}M_\odot$ (open), as a function of redshift. Pink symbols show the observed number densities of quenched galaxies with $M_*> 3\times10^{10}M_\odot$ found by \citet{long:2023} (filled fly) and \citet{carnall:2023} (open diamonds). The number density reported for IllustrisTNG-300 by \citet{hartley:2023} is shown as filled dark blue diamond.
    }
  {\label{fig:numbdens}}
\end{center}
\end{figure}

We define galaxies as ``quenched'' if their current star-formation rate is in fact zero. Using the alternative definition from the literature\footnote{This is the method that was used for example by \citet{remus:2023} to predict quiescent fractions for different halo masses through cosmic time up to $z=4.2$.} based on the specific star formation rate sSFR, i.e., galaxies with $\mathrm{sSFR} < 0.3\times t_\mathrm{Hub}$ are defined as quiescent \citep{franx:2008}, this leads to exactly the same sample of quenched galaxies. At $z=3.4$, for our three mass cuts this leads to $109$ quenched galaxies in cut I) of stellar mass larger than $M_*\geq 2\times 10^{10}M_\odot$, and $36$ quenched galaxies of cut II) with $M_*\geq 3\times 10^{10}M_\odot$, being $28$~centrals and $8$~satellites. In our highest mass cut III), we find $6$ quenched galaxies with $M_*\geq 5\times 10^{10}M_\odot$, two of which are satellites and four are centrals.
\begin{figure*}
  \begin{center}
    \includegraphics[width=.45\textwidth]{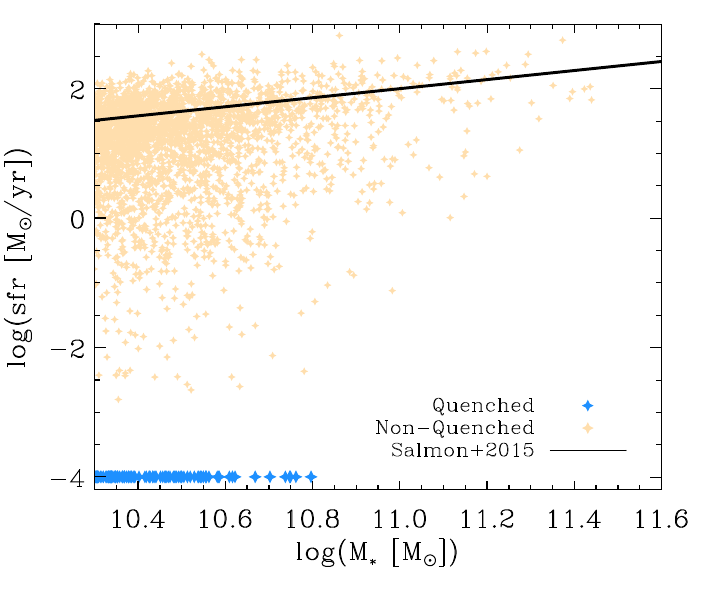}
    \includegraphics[width=.45\textwidth]{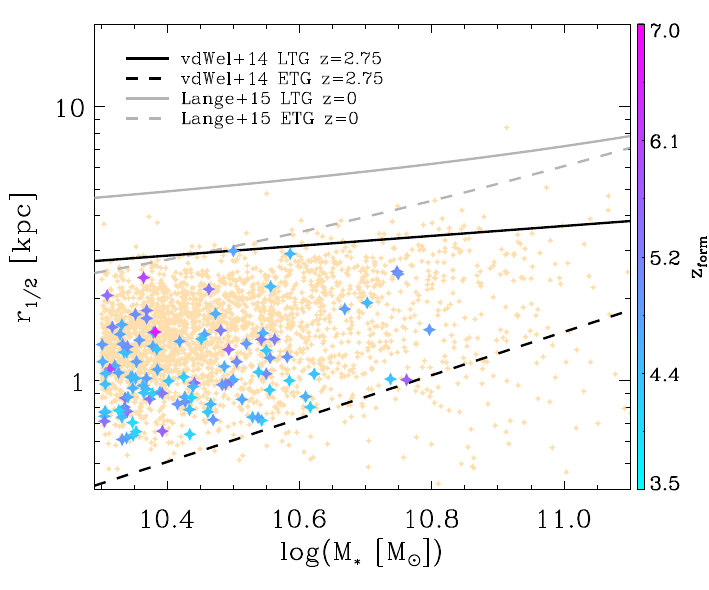}
    \caption{\textit{Left:} The star formation rate (sfr) of Magneticum galaxies at $z=3.42$ as a function of stellar mass. Galaxies with $sfr=0$ are placed at $\log(sfr [\mathrm{M}_\odot/\mathrm{yr}])=-4$ and defined as ``quenched'' (blue). All other galaxies have ongoing star-formation and are defined as ``non-quenched'' (sand). The black line represents the observed star-forming main sequence at $z=4$ from \citet{salmon:2015}.
    \textit{Right:} Stellar mass-size relation of all Magneticum galaxies (sand diamonds) at $z=3.4$, with the quenched galaxies colored according to their formation redshift (see color bar). Black lines show the relations found for early-type (dashed) and late-type (solid) galaxies at $z=2.75$ from \citet{vdwel:2014}, while gray lines mark early-type (dashed) and late-type (solid) relations at $z=0$ from \citet{lange:2015} from the GAMA survey for comparison.
    }
  {\label{fig:main_seq}}
\end{center}
\end{figure*}

Before we study the details of our quenched galaxy sample, we first test if the number densities of quenched galaxies in the simulation is in agreement with observations.
Fig.~\ref{fig:numbdens} shows the number densities of quenched (blue) and non-quenched (sand) galaxies from $z=2$ to $z=10$, for two different stellar mass cuts (open symbols: $M_*>1\times10^{10}M_\odot$, filled symbols: $3\times10^{10}M_\odot$). Generally, the number density of both quenched and all galaxies increases with redshift, but the overall trend flattens at around $z=3$. The increase is stronger for the quenched galaxies, going from around two orders of magnitude less at $z=4.2$ to around one order of magnitude by $z=2$, indicating that more galaxies are quenching their star formation as opposed to those growing massive enough to fulfill the stellar mass cut. Quenched galaxies appear later than non-quenched galaxies of similar mass, with the first quenched galaxies with stellar masses above $M_*>1\times10^{10}M_\odot$ appearing around $z=5.5$. This does not imply that there are no quenched galaxies at higher redshift, but rather that the simulated volume with $128Mpc/h$ box length is still too small to have massive galaxies appear at redshifts as high as $z=8$, as discussed in more detail in the companion paper K23. In addition, there are a few massive galaxies in our simulation that are quenched at higher redshifts, but rejuvenate until the point of measurement at $z=3.4$. One of these galaxies is shown in K23 in more detail, and we will come back to this point later in this study.

Furthermore, Fig.~\ref{fig:numbdens} also provides a comparison to current observed quenched number densities from JWST CEERS (open pink diamonds: \cite{carnall:2023}; pink flys: \cite{long:2023}) using NIRCam imaging and a lower mass cut of about $3\times10^{10}M_\odot$ comparable to the filled blue symbols from the simulation.
We find overall excellent agreement, in particular we see the same increase in the number density when going from a redshift of around $z=4.2$ to $z=3.4$. We do not find as decrease to $z=2.8$ as reported by \citet{long:2023}, however, this could also be due to non-completeness. In addition, we included the quenched galaxy number density reported for the IllustrisTNG-300 simulation by \citet{hartley:2023} at $z\approx4.2$ as dark blue diamond, which is slightly lower than the values found in the observations and the Magneticum simulations. This could be due to the lower resolution of IllustrisTNG-300 compared to the Magneticum simulation used for this study, but also result from different subgrid physics with respect to both AGN and stellar feedback, as both \citet{hartley:2023} and \citet{kurinchi:2023} show for IllustrisTNG quenched galaxies that they are quenched due to the implemented AGN feedback as soon as the kinetic feedback kicks in.
Overall, the similarity in quenched number densities from observations and the Magneticum simulations is promising, so we will now study the properties of the simulated quenched galaxies at a representative redshift of $z=3.4$, the snapshot closest to the redshifts of the observed samples of quiescent galaxies from JWST \citep{long:2023,nanayakkara:2022,carnall:2023} in more detail.

\subsection{Global Properties}
First, we test the general properties of the quenched galaxies compared to the non-quenched galaxies from our simulations and compare with the available observations from the literature. The left panel of Fig.~\ref{fig:main_seq} shows the star formation rate of the galaxies against stellar mass, the so called star formation main sequence, at $z=3.4$ for all galaxies in the simulation (sand color). As can clearly be seen, most of the galaxies lie around the star-forming main sequence as observed by \citet{salmon:2015} at $z\approx 4$ from the CANDELS survey. As expected, there are, however, some galaxies which scatter down towards lower star formation rates. Those galaxies with no ongoing star formation that are therefore classified as quenched are shown in blue (with their logarithmic star formation rate artificially set to -4 so that they can be included in the figure despite their real star formation rate being 0). These are the galaxies that are targeted in this work. This visualizes our selection criterion while also showing that the simulations successfully reproduce the star formation main sequence and thus can be utilized for this study.

As can be seen immediately, none of the quenched galaxies from the simulations is among the most massive galaxies at $z=3.4$. Even more so, most of the massive galaxies are actually on the main sequence, heavily forming stars. This already indicated that the quenched galaxies are, in fact, not hosted by the most massive nodes in the cosmic web at that redshift, but we will return to this in more detail soon. Note that we here already see a difference to the results reported for the IllustrisTNG simulations by \citet{hartley:2023} and \citet{kurinchi:2023}, as in these simulations the quenched galaxies are the most massive ones in the sample.

Before we look at the host halo masses of the quenched galaxies, we study the mass-size relation of the simulated galaxies at $z=3.4$, as shown in the right panel of Fig.~\ref{fig:main_seq}. The simulated galaxies (sand diamonds) clearly lie within the observed range of sizes for a given mass that are found at high redshifts, indicated by the black lines from \citet{vdwel:2014} for galaxies at $z=2.75$, and not within the size range observed at $z=0$ marked in gray \citep[GAMA survey,][]{lange:2015}. As shown already by \citet{schulze:2018}, the Magneticum simulation in the uhr resolution successfully reproduce the mass-size relation of galaxies at $z=0$, and they also reproduce the change in mass-size relation with redshift from $z=2$ to $z=0$ for early-type galaxies \citep{remus:2017}. We clearly see from the right panel of Fig.~\ref{fig:main_seq} that this trend of generally having smaller galaxy sizes for a given stellar mass at higher redshifts continues to even higher redshifts than $z=2$, in good agreement with the observed trends by \citet{vdwel:2014} and also with size evolution trends reported from JWST up to $z\approx4$ \citep{ito:2023}.

Fig.~\ref{fig:main_seq} also reveals that the quenched galaxies (colored diamonds) in our sample on average have smaller sizes at a given stellar mass than the overall sample (sand diamonds), albeit the scatter is similarly large. We do not find a correlation between the stellar mass of the quenched galaxies and their formation redshift, but there is a slight tendency for quenched galaxies that have formed earlier to be larger in size than those that have formed more recently at a fixed stellar mass. While this trend is rather small, it indicates that more recently formed quenched galaxies must have had a larger starburst that leaves them rather compact. 

\subsection{Environment}
\begin{figure}
  \begin{center}
    \includegraphics[width=\columnwidth]{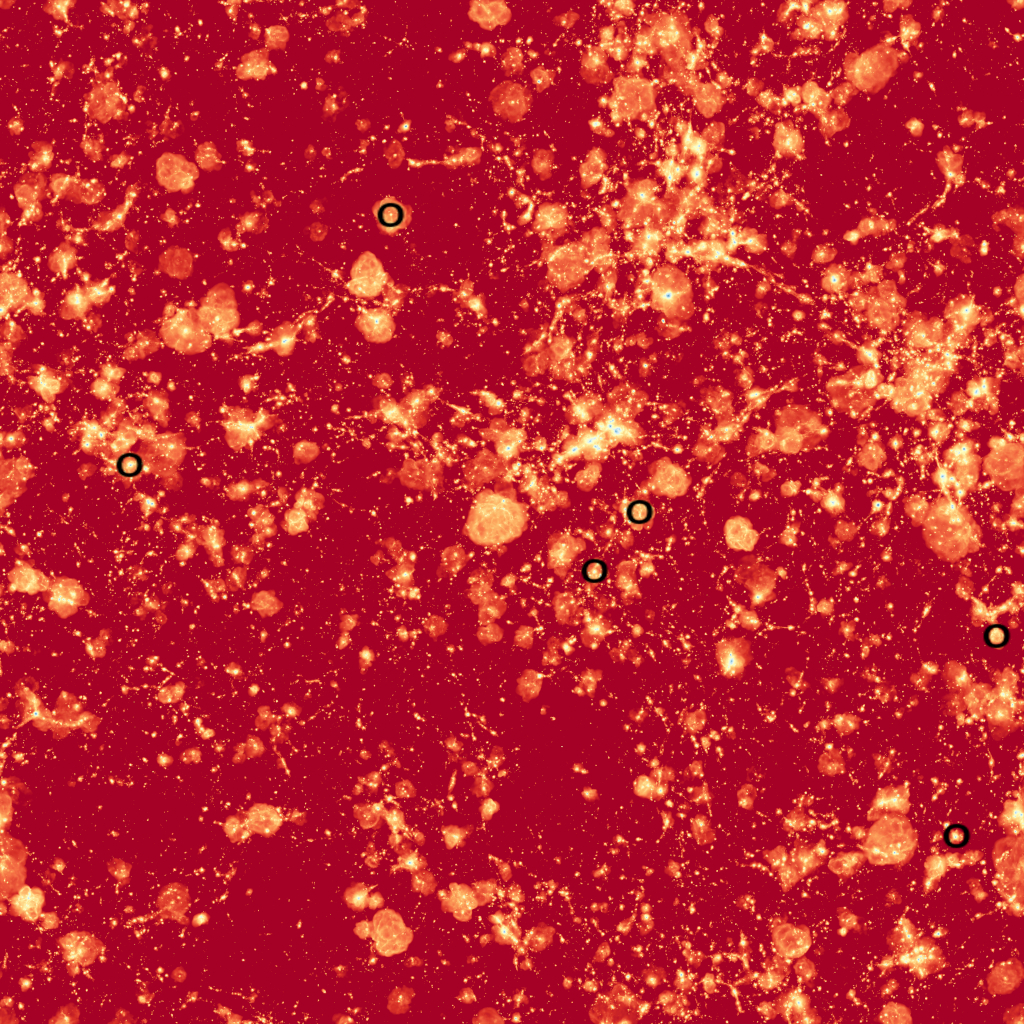}
    \caption{Projected gas temperature of a slice through box4 (uhr) at $z=3.4$. Black circles mark the six most massive quenched galaxies.
    }
  {\label{fig:map}}
\end{center}
\end{figure}
Coming back to the question of the environment of the quenched galaxies, Fig.~\ref{fig:map} depicts the projected gas temperature of a slice through Box3 (uhr) at $z=3.4$. 
The positions of the six most massive quenched galaxies are marked by black circles. All of them exhibit a bubble of heated gas surrounding them, the origin of which is discussed in the companion paper K23 to be the result of their rapid quenching process, driven outward by the feedback from the AGN. Furthermore, they appear to inhabit diverse regions of the cosmic web, though none of them lie at the centers of the hottest regions. Instead, they lie adjacent (left-most and central two), removed (two right-most) or even entirely isolated (top most) from the more pronounced filaments of the cosmic web. We tested this for all $28$ quenched galaxies above a stellar mass of $M_*\leq3\times10^{10}M_\odot$, and found all of them to live in rather remote or adjacent regions of the developing cosmic web, never in the main nodes.
This suggests that quenching at high redshifts is not efficient in the densest environments albeit the AGN feedback is strongest for the most massive galaxies, indicating that the processes that lead to quenching of galaxies at high redshifts are more complicated than the simple picture of AGN feedback quenching the galaxy.

\begin{figure*}
  \begin{center}
    \includegraphics[width=.8\textwidth]{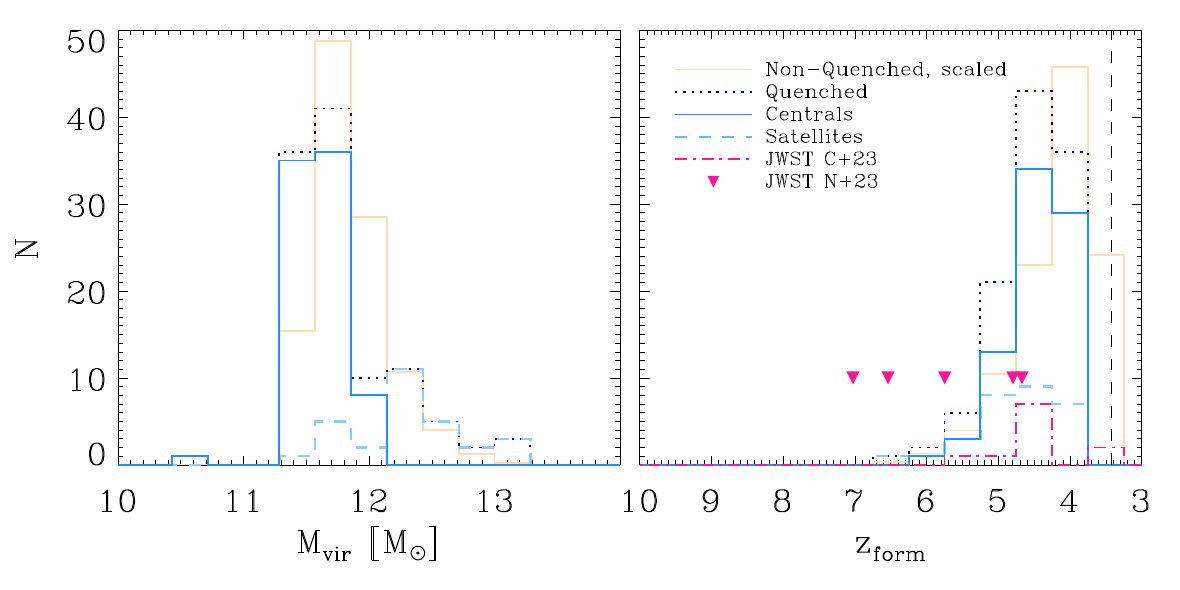}
    \caption{
    \textit{Left:} Histogram of the virial masses of the quenched (blue dotted) and non-quenched galaxies (sand solid line, scaled down to match) with stellar masses above $2\times 10^{10}M_\odot$. The solid blue and dashed blue lines denote quenched galaxies which are centrals and satellites, respectively. 
    \textit{Right:} Histogram of the formation redshifts $\mathrm{z_{form}}$, colored as in the left panel. In addition, observations from JWST are included in pink: the dash-dotted line marks the formation redshift distribution of JWST CEERS quenched galaxies observed between $z=4$ and $z=3$ from \citet{carnall:2023}. The solid triangles mark the formation redshifts of the five galaxies from \citet{nanayakkara:2022} with spectroscopic NIRSpec data.
    }
  {\label{fig:ident}}
\end{center}
\end{figure*}
This is in agreement with what we had seen before already, namely that the quenched galaxies are not among the most massive galaxies. The left panel of Fig.~\ref{fig:ident} further strengthens this point: Shown is a histogram of the virial masses of the quenched (dark blue, dotted) and all (sand, solid) galaxies with stellar masses above $M_*>2\times10^{10}M_\odot$, albeit the latter is scaled to match the numbers of the quenched galaxies for better comparison of the distributions. On first glance, both distributions appear to be similar, with a tendency for the quenched galaxies to be overly common at small virial masses around $M_\mathrm{vir}\approx 5\times10^{11}M_\odot$, that is around the Milky-Way mass range.

However, this picture changes once we split the quenched sample into central galaxies (solid blue line) and satellite galaxies (dashed light blue line). The quenched sample clearly divides in halo mass, with all quenched galaxies in halos of virial masses above $M_\mathrm{vir}\approx 2\times10^{12}M_\odot$ being satellites, i.e. they were accreted onto a more massive dark matter halo, and they are not the central galaxies of the massive nodes. Those quenched galaxies that are central galaxies all live in smaller mass halos, clearly supporting the idea that the environment of a galaxy is crucial for quenching to happen at high redshifts.

\subsection{Formation redshifts - when do the quenched galaxies form?}

Given that the quenched galaxie are already rather massive at $z=3.4$, the question arises when they had actually formed their stars, and whether or not they already differ from non-quenched massive galaxies with respect to their formation times. To this end we define $\mathrm{z_{form}}$ here as the redshift at which half of the galaxies total stellar mass at $z=3.4$ has been formed. 
The right panel of Fig.~\ref{fig:ident}, shows a histogram of the formation redshifts of all galaxies (sand) and the quenched galaxies (blue). As can be seen most galaxies have formed their stars around a redshift of $\mathrm{z_{form}}\approx4$, though there is some scatter towards formation redshifts as high as $\mathrm{z_{form}}\approx7$. The quenched galaxies, both centrals and satellites, have on average formed earlier than the non-quenched galaxies, with the satellites having even earlier formation redshifts than the centrals. These high formation redshifts found for the quenched galaxies are in broad agreement with those found in observed quenched galaxies by \citet[][magenta triangles]{nanayakkara:2022} and \citet[][magenta line]{carnall:2023}. It should be noted, however, that the only quenched galaxy which matches the earliest two times of $\mathrm{z_{form}}\approx7$ found by \citet{nanayakkara:2022} ends up being a satellite galaxy at $z=3.4$.

Nevertheless, we do not have a quenched galaxy with a formation redshift larger than $z=7$, and thus no counterpart for the quenched galaxy reported by \citet{glazebrook:2023} to exhibit a formation redshift around $z=11$. However, earlier formation redshifts can be found in our sample of galaxies if we identify quenched galaxies at $z\approx5$. Observed at that redshift, we even have a quenched galaxy with a formation redshift before $z=8$, as shown by K23. Interestingly, that galaxy rekindles its star formation at about $z\approx4$, and is no longer quiescent at $z=3.4$.

\begin{figure}
  \begin{center}
    \includegraphics[width=.45\textwidth]{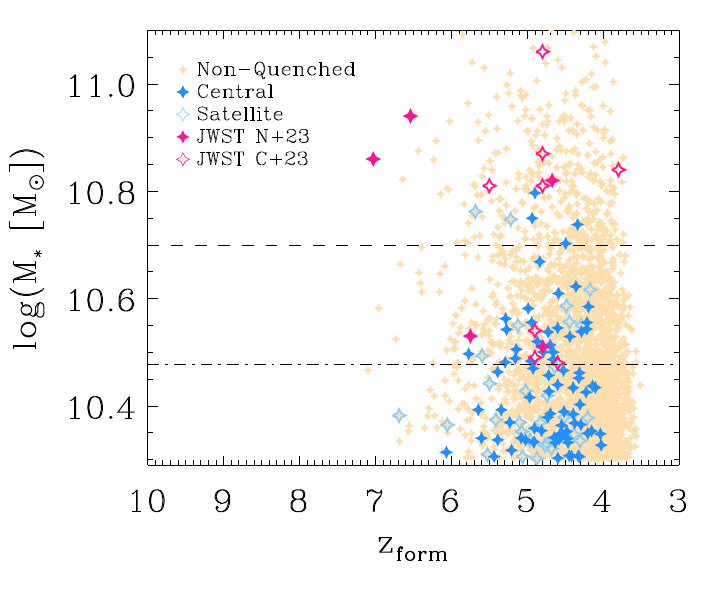}
    \caption{Stellar mass as a function of the formation redshift for the star forming (sand) and quenched galaxies (blue stars), with the latter further differentiated between centrals (filled) and satellites (open) defined at redshift $z=3.4$. Also plotted are observed quenched galaxies from \citet{nanayakkara:2022} and \citet{carnall:2023} (filled and open pink symbols, respectively). The black horizontal lines indicate stellar masses of $5$ (dashed) and $3\times10^{10}M_\odot$ (dash dotted).}
  {\label{fig:props}}
\end{center}
\end{figure}

All observed quenched galaxies at around $z=3.4$ are rather massive in stellar mass, so the question arises if the formation redshift is connected to the stellar mass. Fig.~\ref{fig:props} shows the stellar mass against the formation redshift for the quenched (blue) and non-quenched galaxies (sand) for the simulations, including the observed quenched galaxies (magenta). There is no apparent correlation between the stellar mass either of the quenched or non-quenched galaxies, neither in the simulations nor the observations.

However, while the simulation includes quenched galaxies with masses and formation redshifts comparable to the observations by \citet{nanayakkara:2022} and \citet{carnall:2023} at the mass range from $3\times10^{10}M_\odot$ to $6\times10^{10}M_\odot$, we also clearly see that we do not recover the extremely massive quenched galaxies that are part of the observed sample. The most extreme stellar masses around $10^{11}M_\odot$ are found only for galaxies which are actively forming stars. There are, however, six quenched galaxies with masses which are in excess of $5\times10^{10}M_\odot$, with ids as $ID4180$, $ID5058$ (for the two satellites) and $ID16396$, $ID18258$, $ID20633$ and $ID22007$ (for the four centrals).

\begin{figure}
  \begin{center}
    \includegraphics[width=.9\columnwidth]{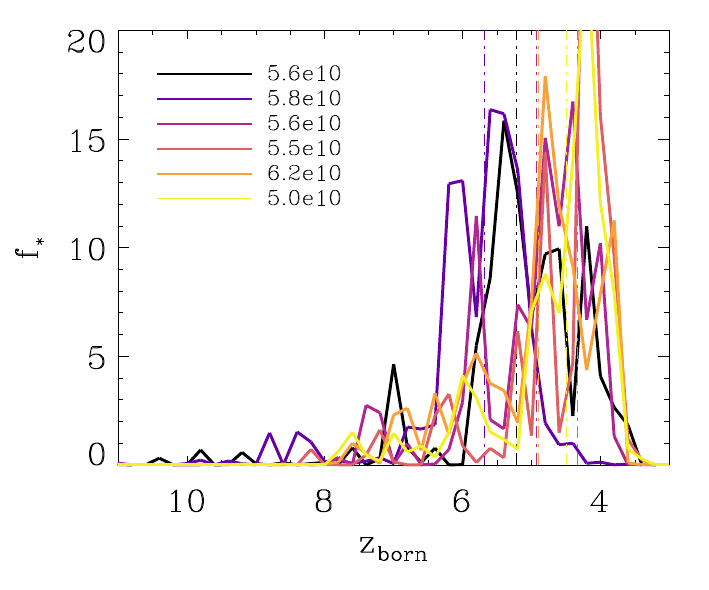}
    \caption{The fraction of stars formed at a given redshift for each of the six most massive quenched galaxies. Vertical colored dash-dotted lines denote their $\mathrm{z_{form}}$, while the label denotes their stellar mass at $z=3.4$ (sorted in order of their ids).
    }
  {\label{fig:age}}
\end{center}
\end{figure}

Fig.~\ref{fig:age} shows the fraction of stars which are formed at a given redshift for these six most massive quenched galaxies (solid lines) as well as their formation redshifts $\mathrm{z_{form}}$ (vertical dash-dotted lines). They exhibit diverse stellar formation histories, with galaxy $ID5058$ (purple) in particular having formed most of its stars very early on. This galaxy is also quenched at an earlier time, with only around $5\%$ of its stars formed after $z=5$ and practically none after $z=4$. Although it is a satellite at $z=3.4$, it was still a central at $z=4.2$ and thus the quenching did not occur due to the environment of the host. In fact, it is quenched in a similar process like the other central galaxies and afterwards accreted by the more massive structure that also prevents any kind of rejuvenation. 

By contrast, galaxies $ID18258$ and $ID22007$ (salmon and yellow) exhibit a strong starburst right around $z=4$, resulting in a much more recent $\mathrm{z_{form}}$. The former in particular has an interesting stellar build up, with three distinct peaks of star formation in sequence at $z\approx 5.2$, $4.8$ and then the final at $z=4$. 

All six galaxies show generally bursty star formation, with significant fractions ($>10\%$) of their final stellar mass formed in these short peaks, usually then followed by dips. As these are the binned formation redshifts of individual stellar particles, this burstiness is not a result of a lack of temporal resolution of snapshot outputs but rather physical in origin. Strong star formation results in higher stellar feedback, which in turn can then briefly inhibit the formation of new stars thus causing a dip. 

It is interesting to note that although this interplay of star formation and feedback had been active in all six galaxies from around $z=6$, they retain significant star formation also during the dips and it is not until close to $z=3.4$ that they fully quench. The final mechanism required to result in a full stop of their star formation is the feedback from the central supermassive black hole, in combination with their environment, which is discussed in more detail in the companion paper K23.
\begin{figure}
  \begin{center}
    \includegraphics[width=.9\columnwidth]{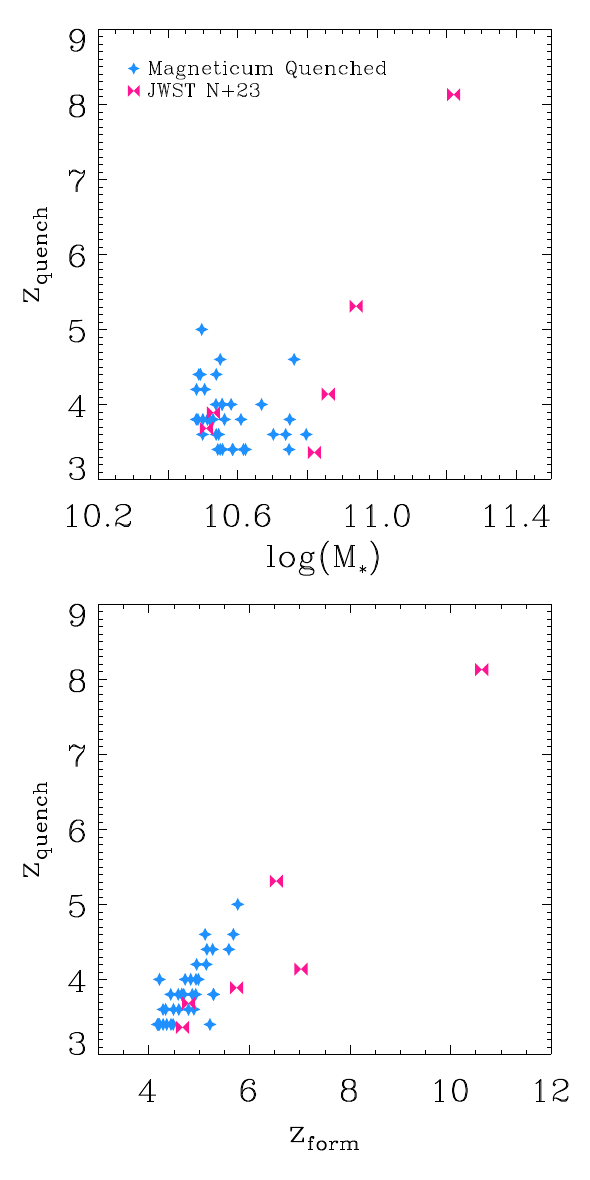}
    \caption{$\mathrm{z_{quench}}$ versus $M_\mathrm{*}$ (\textit{top}) and $\mathrm{z_{form}}$ (\textit{bottom}) for all simulated quenched galaxies (blue), as well as the values of observed galaxies from \citet{nanayakkara:2022} (pink). 
    }
  {\label{fig:quench}}
\end{center}
\end{figure}

The time of quenching, $\mathrm{z_{quench}}$, is defined as the redshift at which the last star in the galaxy as seen at $z=3.4$ has formed. The top panel in Fig.~\ref{fig:quench} shows $\mathrm{z_{quench}}$ as a function of $M_\mathrm{*}$ for all $36$ quenched galaxies with $M_\mathrm{*}>3\times10^{10}M_\odot$ (blue symbols). We find $\mathrm{z_{quench}}$ between $z=3$ to $z=5$, similar to the observed values of \citet{nanayakkara:2022} aside from their most striking outlier. There is no clear additional trend with stellar mass visible, neither in the simulation nor the observations, indicating that the process that quenches galaxies at high redshifts in fact is not directly correlated to the galaxy mass but rather other effects come to play. As discussed by K23, these are a combination of the feedback from the stars, the AGN feedback, and the environment being underdense; this is different than what has been found by \citet{hartley:2023} and \citet{kurinchi:2023} for IllustrisTNG, where the quenching is correlated with the mass of the galaxies due to the implemented feedback. 

Interestingly, the formation redshift $\mathrm{z_{form}}$ and the quenching redshift $\mathrm{z_{quench}}$ are correlated. Galaxies with younger formation redshifts also have lower quenching redshifts, meaning that more recent quenching also implies that half the total stellar mass was formed more recently. This is in agreement with the result that the quenching of the galaxies is preceded by a spike in star formation that forms most of the stars in the galaxy. And while the simulation does not recover the high formation and quenching redshifts of the most extreme outlier from the observations by \citet{nanayakkara:2022} (which is also the galaxy discussed in more detail by \citet{glazebrook:2023}), we clearly see that it lies around where one would expect from the simulated sample when extending the correlation to higher formation redshifts. 

This clearly indicates that the quenched galaxies form most of their stars both in simulations and observations rather quickly in a short time period prior to the very quick quenching event, in agreement with what is discussed by K23. It also shows that this process of quenching must be similar also at higher redshifts, as the behavior of the galaxy from \citet{glazebrook:2023} agrees well with the other galaxies. A larger simulation volume of the same resolution would however be needed to test this hypothesis with respect to the most extreme observed outlier quenched galaxy, which unfortunately is not available at the moment.

\section{What do they become - the multiple pathways of quenched galaxies into the future}\label{sec:future}
As we have shown our simulated quenched galaxies to have similar properties as the observed quenched galaxies at high redshifts, we will now study their evolution forward in time to see if they stay quenched or rekindle their star formation.
In the following we consider the evolution of the sample of galaxies, both quenched and star forming.
As the simulation only evolved in the high resolution down to $z\approx2$, we can only trace the individual galaxy properties down to this redshift, but the overall properties can be followed in lower resolution down to $z=0$.

\subsection{Evolution to Redshift of z=2}
\begin{figure}
  \begin{center}
    \includegraphics[width=.9\columnwidth]{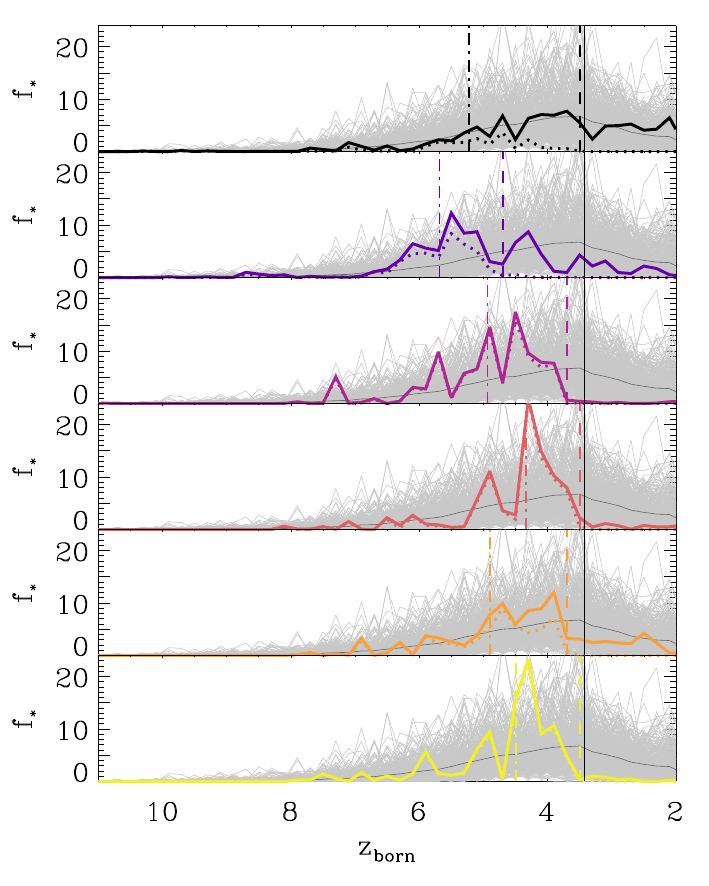}
    \caption{The fraction of stars within the galaxy at $z=2$ as a function of their formation redshift for each of the six most massive quenched galaxies (plotted in order of their ids from top to bottom, $ID4180$, $ID5058$, $ID16396$, $ID18258$, $ID20633$ and $ID22007$). Plotted as colored dotted lines are the fractions at $z=3.4$ when the galaxies are identified as quenched, as shown in Fig.~\ref{fig:age}, re-scaled to the total stellar mass at $z=2$. The light gray lines show the formation histories of all galaxies with $M_*>5\times10^{10}M_\odot$, with their mean in dark gray. The black vertical line denotes $z=3.4$, with the colored vertical lines denoting $\mathrm{z_{form}}$ (dash-dotted) and $\mathrm{z_{quench}}$ (dashed) when observed at $z=3.4$.
    }
  {\label{fig:rejuv}}
\end{center}
\end{figure}
For the six most massive galaxies Fig.~\ref{fig:rejuv} shows their star formation histories down to $z=2$, going from top to bottom in order of the galaxies ids. For comparison, the dotted lines of the same color show the star formation histories of the same galaxies when observed at $z=3.4$, as shown in Fig.\ref{fig:age}, with all dotted lines reaching zero star formation before $z=3.4$.

We find that for three of the central galaxies ($ID16396$ in pink, $ID18258$ in salmon, and $ID22007$ in yellow) there is practically no new stellar mass present at $z=2$, neither accreted nor formed. That is to say they remain quenched down to $z=2$. However, the fourth central quenched galaxy ($ID20633$ in orange) gained stellar mass, as there are more stars present with formation times $\mathrm{z_{form}}>3.4$ compared to what was present in the main progenitor (dotted line). It is however not clear from this figure if the galaxy itself has rejuvenated or if it has only accreted matter.

The first of the two satellites ($ID4180$ in black) eventually merges with its central which is still actively forming stars, thereby significantly increasing its total stellar mass (hence the noticeable deviation to the dotted line also for $\mathrm{z_{form}}>3.4$). At $z=2$ the resulting galaxy is actively forming stars. However, this quenched galaxy was not rejuvenated but rather was the smaller member in a merger event and not the main progenitor of the resulting galaxy at $z=2$. Interestingly, the galaxy that has consumed our quenched galaxy shows a star formation history very similar to the main star formation history of all galaxies above $M_*>5\times10^{10}M_\odot$ shown as gray line.

Finally, galaxy $ID5058$ (lilac) first merges with its central, and thus shows very similar behavior to $ID4180$ for $\mathrm{z_{form}}>3.4$. Towards $z=2$, however, the newly formed galaxy shows very little ongoing star formation, in stark contrast to galaxy $ID4180$. This is because after merging with its central, $ID5058$ then falls onto an even larger nearby galaxy group becoming a satellite again at $z=2$. Here, the second galaxy also experiences quenching after it has consumed our quenched galaxy, this time, however, the quenching is a slow process as it is due to the new hosts environment. It is the only case of environmental quenching that we found in our sample of quiescent galaxies at high redshifts.

Compared to the mean of the total sample of galaxies plotted in gray, the three which remain fully quenched naturally deviate most strongly toward earlier formation times. Interestingly, the two satellites which merge however show largely different behavior from one another, with $IS4180$ even exceeding the total sample's average fraction of stars formed at later times $z\approx2$, while $ID5058$ suffers a second quenching event.

\begin{figure*}
  \begin{center}
    \includegraphics[width=.9\textwidth]{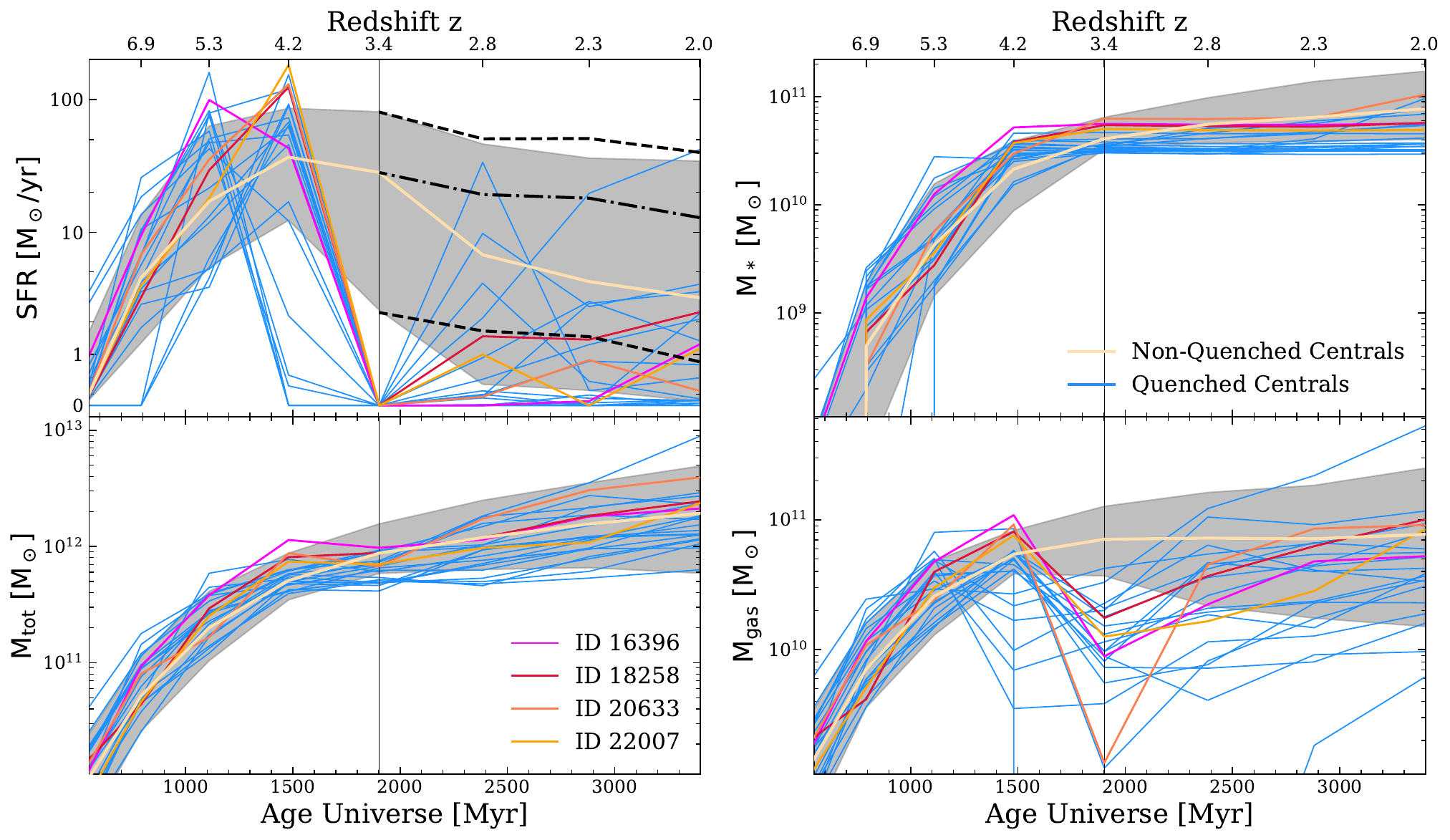}
    \caption{Redshift evolution of the star-formation rate (\textit{upper left}), stellar mass (\textit{upper right}), the total mass (\textit{lower left}) and gas mass (\textit{lower right}) of centrals within the II) mass cut  which remain centrals down to $z=2$. For the $817$~non-quenched galaxies the median is shown as sand color line, with $1\,\sigma$ as gray shaded. The individual evolutionary tracks of the $22$~quenched galaxies are in blue, with the four most massive main galaxies at $z=3.4$ as colored lines as in Fig.~\ref{fig:rejuv}. For the star-formation rate the $y-$axis is scaled with asinh. The black dash dotted line shows the mean SFR of all centrals (also including galaxies which fulfilled the mass cut only after $z=3.4$), with the dashed black lines showing the $1~\sigma$ range of this larger sample. 
    }
  {\label{fig:evo}}
\end{center}
\end{figure*}

To see how other properties of the quenched galaxies evolve with time compared to like galaxies, we consider the broader sample of the II) mass cut. We reduce the sample to galaxies which are centrals at $z=4.2$ and remain centrals until $z=2$ to avoid including galaxies which may have quenched environmentally as opposed to internally. This results in $817$~non-quenched and $22$~quenched centrals, with their evolution of galactic properties shown in Fig.~\ref{fig:evo}.

The top left panel depicts the star formation rate of the sample. We find that the quenched galaxies all spike in their star formation rate in either of the two snapshots ($z=4.2$ or $z=5.3$) immediately before being quenched by $z=3.4$. All but one exception reach star-formation rates in excess of the median of non-quenched centrals (sand color line), with half exceeding the median by more than $1\,\sigma$ at either $z=4.2$ or $z=5.3$ (gray shaded area).
\begin{figure*}
  \begin{center}
    \includegraphics[width=.9\textwidth]{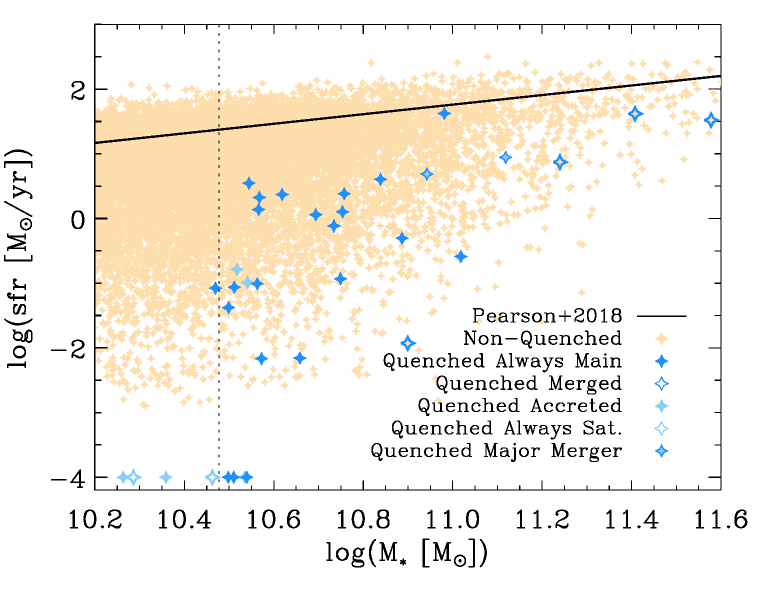}
    \caption{Star formation rate versus stellar mass at $z=2$ of the descendants of our sample of galaxies with mass cut II), with the observed main sequence from \citet{pearson:2018} shown as a black line. Tan symbols are galaxies which were forming stars at $z=3.4$, while blue symbols were quenched. We differentiate the second group into five subgroups, depending on their merger history post-quenching.
    }
  {\label{fig:sfr_evo}}
\end{center}
\end{figure*}

It is interesting to note here that the non-quenched sample of galaxies already more massive than $M_*>3\times10^{10}M_\odot$ (sand line and gray shade) at $z=3.4$ also strongly drops in star formation rate toward $z=2$, with the peak of its median at $z=4.2$. This drop is much stronger than the slight decline found if all centrals with $M_*>3\times10^{10}M_\odot$ at a given redshift are included, as shown by the black lines, which include also those which only recently rose above the stellar mass cut (median black dash-dotted, $1\,\sigma$ bounds black dashed). Consequently, if one were to observe all central galaxies at $z=2$ those which have been massive centrals since $z=4.2$ (around $2\,\mathrm{Gyr}$) will generally have lower star formation rates. This also indicates that those galaxies that newly cross the mass threshold are still evolving due to ongoing star formation more strongly than those galaxies that have already formed stars early on.

The excess in star formation rate at early times correspondingly leads to the quenched galaxies lying at the higher end of stellar mass at $z>4$, as can be seen in the top right panel of Fig.~\ref{fig:evo}. All but two of the $22$~quenched centrals lie above the median stellar mass at $z=4.2$ (sand color line). As all galaxies (quenched and non-quenched) are selected based on their stellar mass at $z=3.4$ this means that the quenched galaxies generally formed their stellar mass earlier compared to like galaxies. Following their evolution forward in time most remain quenched, forming (or accreting) little to no additional stellar mass over a time of around $t\approx1.5\,\mathrm{Gyr}$. However, there are a few which gain sufficient amounts of new stars to reach the median of the total sample. 

Interestingly, the total mass of the galaxies dips below the median for most quenched centrals at $z=3.4$ (lower left panel of Fig.~\ref{fig:evo}). This is largely due to the significant ejection of gas which occurs during quenching (lower right panel of Fig.~\ref{fig:evo}). The gas is then replenished less quickly compared to the accretion of dark matter, resulting in the descendants at $z=2$ of quenched centrals having average to slightly reduced total masses but significantly less gas and stellar mass. This means that the reduction in baryon fraction caused by the quenching process as described by K23 remains in place for most cases up to a time of $1.5\,\mathrm{Gyr}$. 

As we saw in the top left panel of Fig.~\ref{fig:evo}, the quenched centrals do not necessarily remain quenched. To define then whether a quenched galaxy has rejuvenated at some point after $z=3.4$, we must consider what we mean by \textit{rejuvenation}. It is for example possible that the quenched galaxy merges with a larger other galaxy which has ongoing star formation, as noted earlier for the cases of $ID4180$ and $ID5058$. 

We begin first by considering the sample at $z=2$, simply asking whether the quenched galaxies' descendants have rekindled their star formation to become normal main-sequence galaxies. Fig.~\ref{fig:sfr_evo} shows the star forming main sequence at $z=2$, with the blue stars being the traced descendants of the quenched galaxies and sand color points being all non-quenched galaxies (not only traced galaxies). 

The quenched galaxies are split into five groups. First, those which were a satellite at the time of quenching at $z=3.4$ and have remained a satellite down to $z=2$ (``always sat''). Second, those which were centrals at $z=3.4$ but were accreted onto another larger halo but have not yet merged (``accreted''). Third, quenched galaxies which merged with another more massive structure, where the original quenched galaxy cannot be separated anymore (``merged''). Fourth, there are two cases where at the time of quenching the galaxy is closely interacting with another galaxy of comparable mass, and merges by $z=2$ (``major merger''). Finally, those which remained centrals without ever merging with a larger galaxy (``always main''). 

Group one contains three galaxies (open light blue symbols). They all remain fully quenched, and the hot halo of their host galaxies (two of which become groups by $z=2$) is thus efficient enough to suppress any rejuvenation of the star formation within the massive quenched galaxies already at $z>2$. They also experience stripping, as their mass now is below our original stellar mass cut of $M_*=3\times10^{10}M_\odot$, as marked by the dashed black line.  
\begin{figure*}
  \begin{center}
    \includegraphics[width=0.95\textwidth]{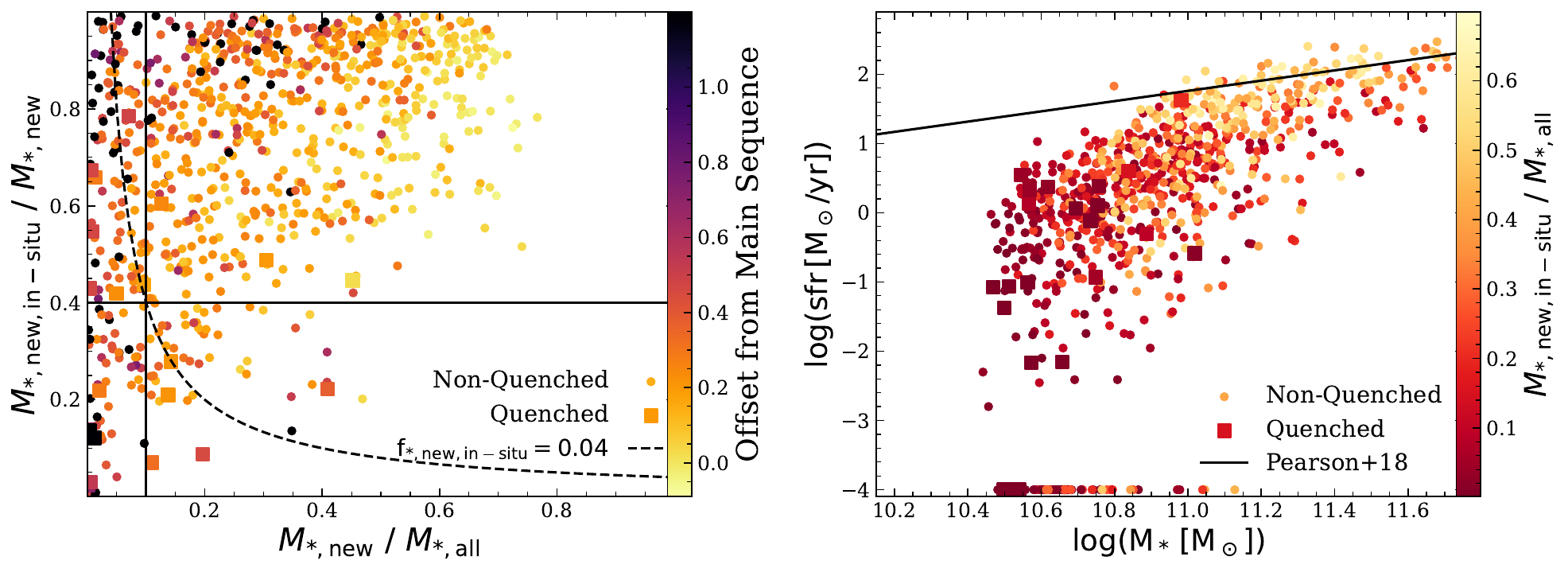}
    \caption{\textit{Left:} Fraction of young stars which are self-made versus fraction of total stars which are young for traced centrals at $z=2$, defined as non-quenched (dots) or quenched (square) based on the SFR at $z=3.4$. Colored by a galaxy's perpendicular offset from the $z=2$ main-sequence, normalized between $0$ (lies on MS, yellow) and $1$ (farthest offset to MS, purple), with starburst galaxies lying above the MS at $<0$ (yellow) and galaxies with $SFR=0$ in black. The black dashed line denotes a fraction of young in-situ stars $f_{*,new,in-situ}=M_\mathrm{*,new,in-situ}/M_\mathrm{*,all}=0.04$. \textit{Right:} As Fig.~\ref{fig:sfr_evo}, but only galaxies which remained centrals from $z=3.4$ to $z=2$. Colored by the fraction of total stellar mass contained in young in-situ stars. 
    }
  {\label{fig:cumuage}}
\end{center}
\end{figure*}

There are five quenched galaxies which are accreted onto other structures post-quenching (filled light blue symbols). Their star formation rates are very low, with two fully quenched. However, unlike the ``always sats'', they are not entirely shut-down with three retaining very low residual star formation that they rekindled prior to infall in the more massive structure. Here, we see environmental quenching at work as these galaxies slowly loose their star formation abilities. They highlight that even for how massive the quenched galaxies are at $z=3.4$, they do not necessarily remain the locally dominant node. 

Four quenched galaxies merge (open dark blue symbols) with a more massive system to become a central by $z=2$ (including~$ID4180$). They contribute between $1/10$ to $1/3$ of the stellar mass at $z=2$, and the descendant galaxy has high mass and star forming activity. These are curious cases then where a quiescent galaxy is assembled onto a more massive star forming galaxy, and the star formation is not quenched, opposite to the common pathway found at low redshift. Such remnant galaxies have extensive stellar bulges and are closer to S0 galaxies than typical disk galaxies.

Then we have two galaxies undergoing a major merger around the time of quenching (two-color symbols), one of which is~$ID5058$. Both show a large growth in stellar mass and high star forming activity, being very comparable to the behavior seen for the ``merged'' category. It follows that so long as the galaxy with which the quenched one merges is of equal or greater mass, the odds are it will become a ``normal'' sub-main sequence galaxy. 

The final group are the $22$~quenched centrals which remained centrals and never merged with any larger galaxy down to $z=2$ (filled dark blue symbols). They represent the largest group, as well as the one with the most scatter in resulting star formation rates. These are the ones which we are most interested in for considering how many galaxies can rejuvenate from internal quenching through gradual feeding from the cosmic web. 

Of these~$22$, we find a few which lie near the main sequence, with one in particular fully rejuvenating to lie on the relation. The majority lie at middling to low star formation rates, while three are still fully quenched. None have grown to stellar masses $M_*>1\times10^{11}M_\odot$, with about half growing by less than $10\%$. Nonetheless, we find that star formation is not permanently shut off even for central galaxies which undergo violent quenching, eject their gas and which subsequently have a large SMBH at their core. These are not rejuvenated via major mergers bringing in significant gas, or by merging themselves with a more massive starforming galaxy, but rather through accretion of gas from the cosmic web, and we will study this process in more detail in the following.

\subsection{Rejuvenation}
We differentiate between stars formed in other galaxies and are then accreted versus those which are formed within the galaxy itself, so in-situ. To consider a newly formed star as in-situ made, we require the progenitor gas particle to be within the virial radius of the central galaxy, and the resulting star to be bound to the central immediately afterwards. This gives four categories of stars in a given galaxy at $z=2$: the initial old component which was in the galaxy at $z=3.4$ (``Old, In Halo''), old stars which were formed elsewhere and accreted sometime after $z=3.4$ (``Old, Accreted''), and younger stars (formed after $z=3.4$) either made in-situ (``New, In-Situ'') or accreted (``New, accreted''). 

In the left panel of Fig.~\ref{fig:cumuage} we show the mass fraction of in-situ formed younger stars, $M_\mathrm{*,new,in-situ}/M_\mathrm{*,new}$ as a function of the fraction of total stellar mass contained in younger stars, $M_\mathrm{*,new}/M_\mathrm{*,all}$ for central galaxies at $z=2$ traced from $z=3.4$. We find that galaxies which have formed a significant fraction of their final mass recently tend also to have formed most of that mass in-situ. This is because high-redshift galaxies contain significant fractions of gas, such that any accretion event involving young stars will also involve bringing in a lot of gas, enabling in-situ star formation. Consequently, there emerges a lower bound to the fraction of young stars made in-situ with a slope of around~$1$. Put another way, if $x\%$~of all stars in a traced galaxy at $z=2$ are young ($<1.5\,$Gyr old), then at least $x\%$~of these young stars were made in-situ (though the range can go up to $100\%$ made in-situ for any given $x$). 
\begin{figure*}
  \begin{center}
    \includegraphics[width=0.975\textwidth]{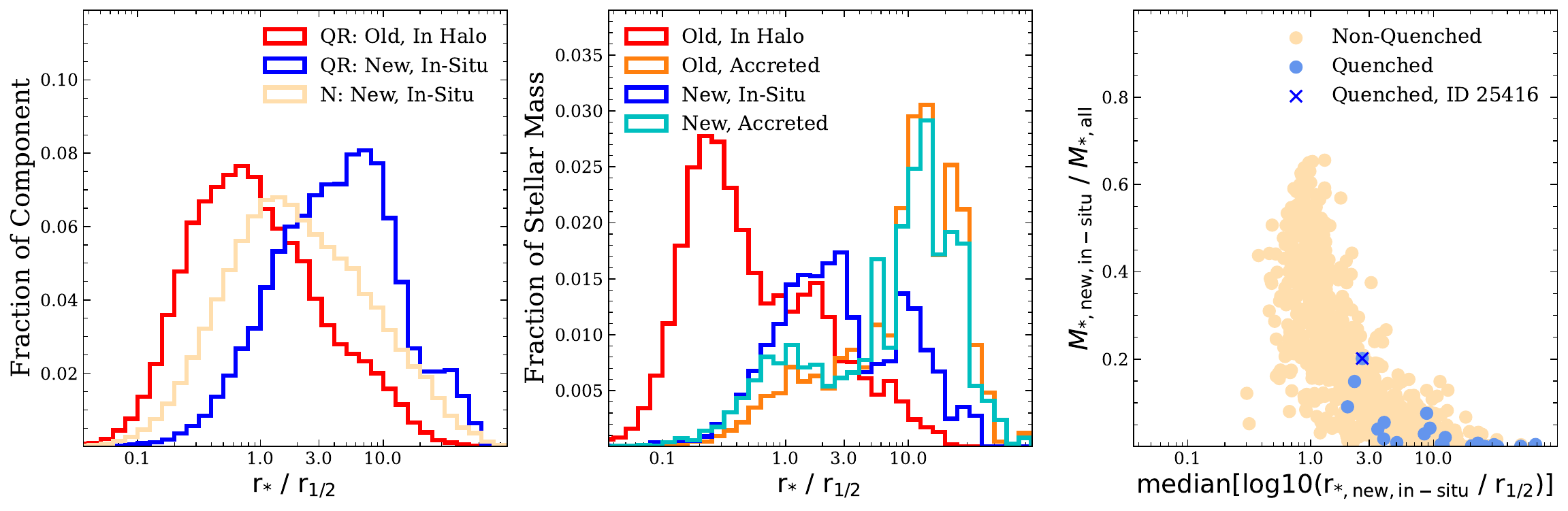}
    \caption{\textit{Left:} Radial distributions of the young in-situ formed stars for the descendants of rejuvenated quenched (blue) and non-quenched (sand) galaxies at $z=2$. The old component which was in the halo prior is also shown for the quenched descendants (red). \textit{Middle:} The radial total mass fraction of different components (legend) for the quenched galaxy~$ID25416$ which rejuvenates the strongest. \textit{Right:} Fraction of total stellar mass which is young and in-situ made versus the its median radial position. All radii are given relative to the stellar half mass radius. 
    }
  {\label{fig:cumurad}}
\end{center}
\end{figure*}

For the quenched galaxies this trend is largely absent, as there are multiple quenched galaxies with many young stars but few of which are in-situ made. This indicates that galaxies which have quenched previously are better at prohibiting rekindling of their own star-formation even when they accrete significant amounts of new matter compared to other non-quenched galaxies. 

However, we also find that prior quenching of a galaxy does not mean that it cannot rejuvenate. Indeed, there are six galaxies which post-quenching end up making at least $4\%$ of their total stellar mass in-situ afterwards. Three of them simultaneously grow by more than $10\%$~while making at least $40\%$~of these new young stars in-situ (lie in the upper right quadrant). Nine exhibit some lesser degree of star-forming activity, with five in the upper left (little new stars but mostly made in-situ) and four in the lower right (many new stars but few of them self made), while seven remain effectively quenched (lower left). This means we find that around $30\%$~rejuvenate, $40\%$~rekindle some residual star forming activity and $30\%$~remain quenched throughout. 

The location in the $M_\mathrm{*,new,in-situ}/M_\mathrm{*,new} - M_\mathrm{*,new}/M_\mathrm{*,all}$ plane gives the \textit{integrated} star-forming activity, but how does it compare to the \textit{instantaneous} star forming rate? To see this we color the points by their perpendicular distance to the star-forming main sequence at $z=2$ as given by \citet{pearson:2018}, normalized from $0$ (on the MS, dark yellow) to $1$ (far below, dark purple), with $>1$ being quenched (black) and $<0$ being star-bursting galaxies above the MS (yellow). 

Interestingly, galaxies which end as quenched at $z=2$ (black points) preferentially stick towards the perimeter, having either very little young stars (lie far left) or having formed practically all younger stars in-situ (lie at the top). We interpret this as them quenching either through starvation caused by lack of accretion (low $M_\mathrm{*,new}/M_\mathrm{*,all}$) or through the rapid starburst quenching (RSQ) mechanism described by K23, with a collapse of significant gas followed by a massive starburst (high $M_\mathrm{*,new,in-situ}/M_\mathrm{*,new}$) and combined stellar and AGN feedback which quenches the star formation.

Generally, we find that when comparing galaxies which made a similar fraction of new stars in-situ, those which show a higher instantaneous SFR (closer to main sequence, more yellow) lie farther right (have a higher total fraction of younger stars). This also implies a higher total fraction of new in-situ formation (an \textit{integrated} quantity) based on the current SFR (an \textit{instantaneous} quantity). To validate this we plot the star formation rate versus stellar mass in the right panel of Fig.~\ref{fig:cumuage} and color by the total fraction of young in-situ formed stars $M_\mathrm{*,new,in-situ}/M_\mathrm{*,all}$. We indeed find that galaxies lying closer to the MS show higher $M_\mathrm{*,new,in-situ}/M_\mathrm{*,all}$, though there is some scatter which warrants a closer look in the future. 

Having seen that around~$30\%$ of the quenched galaxies rejuvenate we consider now where they make these new in-situ stars, in particular when compared to galaxies which were not quenched at $z=3.42$. We find in the left panel of Fig.~\ref{fig:cumurad} that the new in-situ stars in the six quenched descendants (blue color) which rejuvenated significantly ($M_\mathrm{*,new,in-situ}/M_\mathrm{*,all}>0.4$) lie at noticeably farther radii than in the non-quenched galaxies (sand color), indicating that the most central region does not experience significantly rekindled star formation once a galaxy has been quenched. This may also in part be caused by the comparatively more compact central bulges formed from the starburst which precedes the quenching (see also Fig.~\ref{fig:main_seq}), where at $z=2$ the non-quenched sample have $\bar{r_{1/2}}\approx3.1\mathrm{kpc}$ while the quenched have $\bar{r_{1/2}}\approx1.8\mathrm{kpc}$. 

Taking the descendant of the quenched galaxy which has rejuvenated the most, galaxy~$ID25416$, we plot the origin of its stellar component in the middle panel of Fig.~\ref{fig:cumurad}. The most central region 
within $1\, r_\mathrm{1/2}$ is dominated by the original stars formed in the burst prior to quenching (red line). The stars which are formed in-situ post quenching primarily reside between~$1\, r_\mathrm{1/2}$ and~$3\, r_\mathrm{1/2}$, where they also are the dominant component. Beyond around~$5\, r_\mathrm{1/2}$ the accreted components begin dominating, with comparable mass contained in younger versus older stars.

\begin{figure}
  \begin{center}
    \includegraphics[width=0.95\columnwidth]{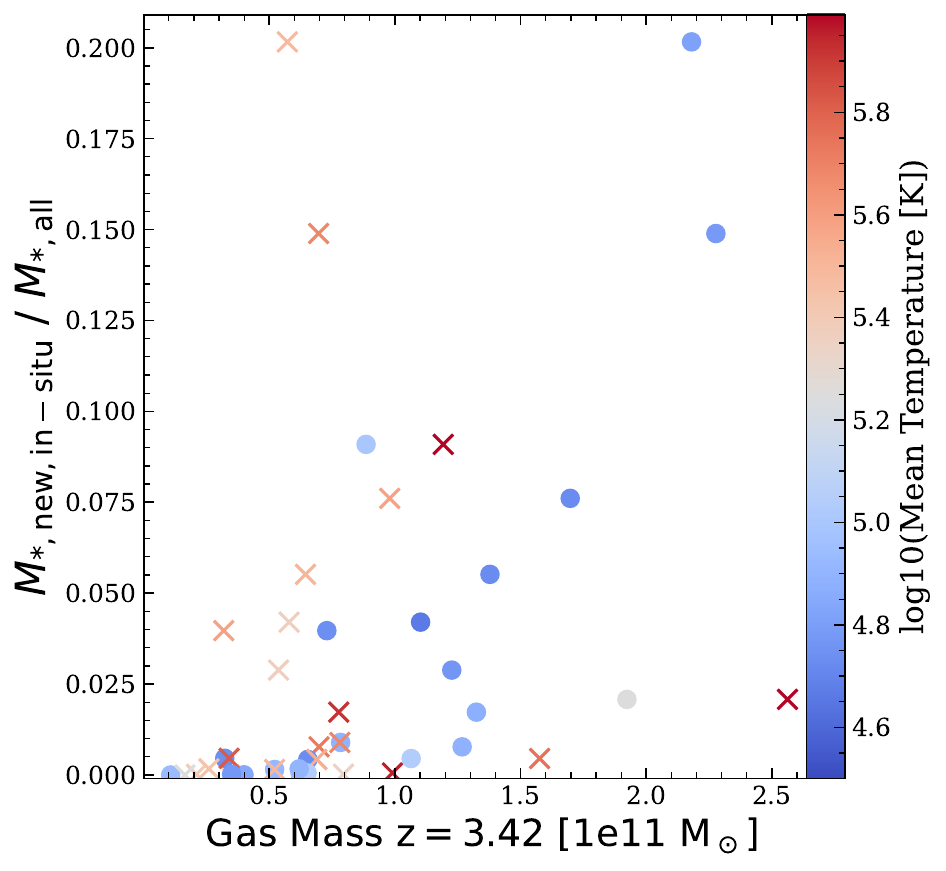}
    \caption{
    The fraction of stellar mass at $z=2$ which was formed in-situ post-quenching (after $z=3.4$) versus the gas mass within three virial radii around the galaxy immediately following quenching (at $z=3.4$). Gas is split into inflowing (dots) and outflowing components (crosses), and colored by the mean temperature.}
  {\label{fig:rejuv_env}}
\end{center}
\end{figure}

Finally we show the fraction of total stellar mass in young in-situ stars as a function of their median radius for all individual galaxies. We find that galaxies with little recent in-situ star formation also tend to have this formation occurring farther out. As the fraction of in-situ formed stars increases they begin to dominate the total stellar mass budget, until eventually the stellar half mass radius coincides with their median radius. 

We then ask whether the degree of rejuvenation can be predicted from observables at the time of quenching already. Fig.~\ref{fig:rejuv_env} shows the fraction of young in-situ formed stars versus the surrounding gas within three virial radii, split by outflowing and inflowing components defined by $\vec{v}\cdot\vec{r}>0$ and $<0$ (crosses and dots, respectively). The amount of mass is comparable in both components, and we find the outflowing gas to be generally hot ($>10^{5.4}\mathrm{K}$) while the inflowing gas is colder. We further see that a larger amount of inflowing cold gas (blue) results in a higher amount of rekindled in-situ star formation, while there is no such correlation with the amount of outflow. 

This dependence of rejuvenation on the inflowing gas mass indicates that it is the environment which ultimately determines the fate of high-z quenched galaxies as opposed to for example the mass of their AGN. Indeed, we find no correlation between the amount of rejuvenation $M_\mathrm{*,new,in-situ}/M_\mathrm{*,all}$ and the relative mass fraction of the black hole $M_\mathrm{BH}/M_\mathrm{*}$ at $z=3.4$ for the quenched galaxies. This mirrors the results by K23, where the long-term evolution of the star formation is most tightly correlated with the environment.

\subsection{Descendant Galaxies at z=0}
Finally, we want to see where the quenched galaxies end up at present day. As the higher resolution Box3 uhr ran until $z\approx 2$, we match at the last snapshot the galaxies to their counterparts in the lower resolution version of the same box, Box3 hr, which ran until $z=0$. Note that exact matches are difficult in the case of satellites. However, as we care only about the final halo in which they end up in, it matters only that we can match the correct larger halo, which is generally possible. 

\begin{figure*}
  \begin{center}
    \includegraphics[width=1.9\columnwidth]{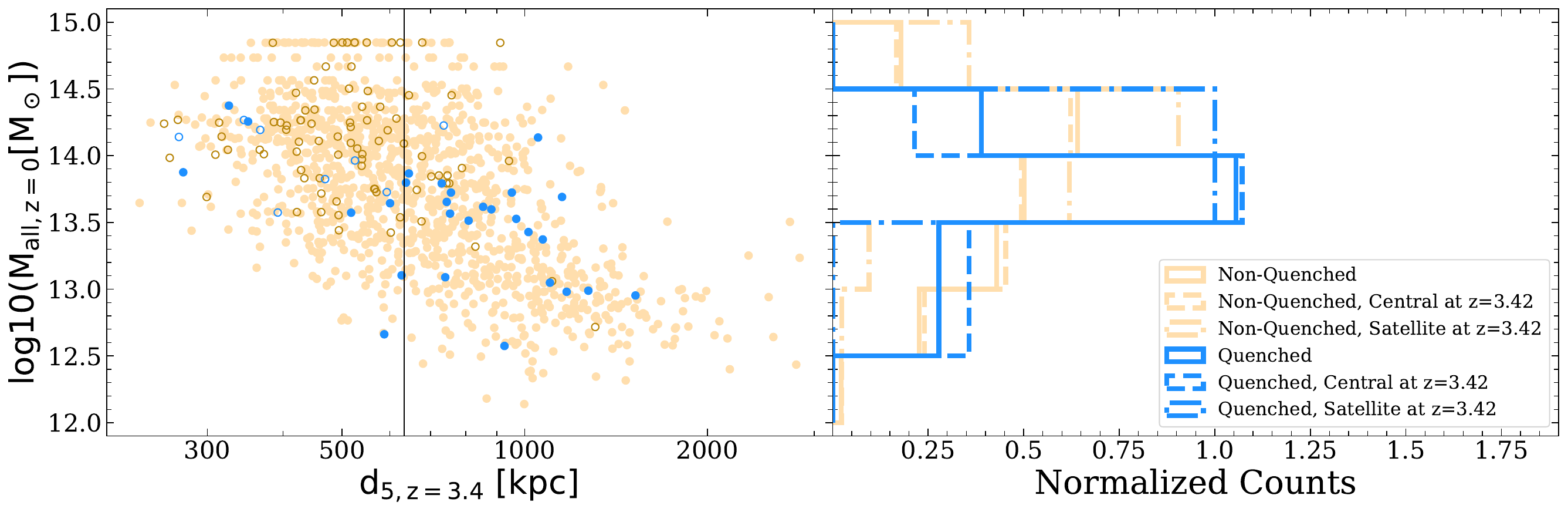}
    \caption{\textit{Left:} Total halo mass at $z=0$ plotted against a tracer for the environment at $z=3.4$, $d_5$, for quenched (blue) and non-quenched (sand) centrals (filled symbols) and satellites (open symbols), defined at $z=3.4$. The median $d_5$ for all galaxies is given by the vertical black line. \textit{Right:} Binned total halo mass at $z=0$. 
    }
  {\label{fig:finmass}}
\end{center}
\end{figure*}

The right panel of Fig.~\ref{fig:finmass} shows the histogram of the total mass of the host halo $M_\mathrm{all}$ at $z=0$ for the descendants of the quenched (blue lines) and the non-quenched galaxies (tan lines). 
We find that the quenched galaxies generally end up in galaxy groups of a total mass around $M_\mathrm{all,z=0}\approx 10^{13.5}M_\odot$. This is particularly true for the galaxies which were centrals at $z=3.4$ (dashed blue line), with only three of $28$ ending in galaxy clusters with $M_\mathrm{all,z=0}\geq 10^{14}M_\odot$. 

This means that on average they end in less massive halos compared to the total sample. The median mass of the final halo for non-quenched galaxies (sand lines) is $6.6\times10^{13}M_\odot$, while for quenched galaxies it is $4.4\times10^{13}M_\odot$. This difference is even more significant for centrals, with $5.8\times10^{13}M_\odot$ versus $3.8\times10^{13}M_\odot$, while for satellites the final halos are similar ($1.3\times10^{14}M_\odot$ versus $1.2\times10^{14}M_\odot$). That is to say we find that a massive central galaxy at $z=3.4$ will on average end up in a halo of $50\%$ higher mass at $z=0$ if it is forming stars versus if it is fully quenched.

This imprint of the eventual halo mass is also present in the environment at higher redshifts. As an environmental tracer we define here $d_5$ as the radius of the sphere required to include five neighboring galaxies with total masses $M_\mathrm{all}\geq10^{11}M_\odot$. The final halo mass at $z=0$ is then plotted against this radius calculated at $z=3.4$ in the left panel of Fig.~\ref{fig:finmass}. 

We find that there is a diffuse correlation between a more dense environment at $z=3.4$ (lower $d_5$) and a higher final halo mass at $z=0$. This is overall to be expected as \citet{remus:2023} find that the environment is the best tracer for the $z=0$ mass of protoclusters. What is striking, however, is that of the $28$~quenched centrals (blue filled dots) $75\%$~lie above the median separation of all galaxies (vertical black line), so lie in underdense environments. 

This may mean that a specific type of environment is required to fully self-quench through combined stellar and AGN feedback. Indeed, it is reasonable to assume that denser environments are more proficient at replenishing lost gas. To remove enough gas to completely cease star formation requires so little gas accretion that the feedback can overpower the inflow. This idea is explored further in K23 and explains why here the quenched galaxies are found predominantly in underdense environments.

\section{Summary and Conclusions}\label{sec:conclusion}
We studied quenched galaxies identified at $z=3.4$ in the Magneticum Pathfinder large high-resolution Box3 uhr and compared their properties to observations of quenched galaxies from JWST by \citet{nanayakkara:2022} and \citet{carnall:2023}. We show that the simulation can successfully reproduce similar quenched fractions for comparable stellar masses to those observed, albeit we cannot reproduce the most massive of the quenched galaxies as the box volume of the simulation is still too small to capture these biggest nodes. 

We find the quenched galaxies to be rather compact in size compared to non-quenched galaxies, in good agreement with recent observations by \citet{ito:2023}. Furthermore, the formation redshifts of the simulated quenched galaxies are in good agreement with the observed formation redshifts, with some reaching up to $z_\mathrm{form}\approx6$, aside from the observation of a peculiar quenched galaxy reported by \citet{glazebrook:2023} which is older than any formation redshift we can produce here. However, our full simulation sample contains at least one galaxy that is quenched at about $z=5$ and which has a formation redshift beyond $z=8$ but which rejuvenated afterwards, as shown in the companion study to this work by K23.

All our quenched galaxies experience a fast period of star formation followed by a rapid decline that leads to quenching, resulting in similar quenching redshifts as observed.
In fact, we find that the quenching redshifts and formation redshifts of the quenched galaxies are correlated, with galaxies with higher formation redshifts also having higher quenching redshifts. This is found for both simulations and observations to be the case, and the outlier quenched galaxy from the observed sample by \citet{glazebrook:2023} actually follows the same trend as the simulations predict if extrapolated to higher redshifts. This indicates that quenching at high redshifts proceeds generally in a similar manner, such high redshifts occurs on a very short timescale after a massive star formation event, in agreement with what is discussed by K23.

We follow the quenched galaxies forward in time, and find that by $z=2$, that is within $1.5\mathrm{Gyr}$ after being fully quenched, about $20\%$ of the galaxies that are centrals at $z=3.4$ are accreted onto a more massive structure by $z=2$, and are either still satellites of that structure or have merged with the more massive central galaxy. Of the remaining $80\%$~of quenched galaxies, $30\%$~rejuvenate, $30\%$~remain fully quenched, and $40\%$~develop some residual star formation. Those galaxies that rejuvenate can even reach the star formation main sequence and appear at a later time to be normal star-forming galaxies.

For those quenched galaxies that rejuvenate, we find the star formation to primarily occur on the outskirts between~$1\, r_\mathrm{1/2}$ and~$3\,r_\mathrm{1/2}$, not reaching the central regions. This is different the non-quenched comparison sample of galaxies that are already massive at $z=3.4$, where star formation can still occur at the center even down to $z=2$. Furthermore, we find that the amount of newly formed stars post-quenching primarily correlates with the amount of cold gas that has been inflowing, while there is no correlation to the outflowing hot gas component, clearly showing that the rejuvenation only depends on new inflowing gas reaching the galaxy, independent of the mass of the AGN that drives the hot wind outwards.

Finally, we find that quenched galaxies at $z=3.4$ tend to end up in less massive final halos at $z=0$ compared to non-quenched massive galaxies at $z=3.4$. This difference is negligible for galaxies that become satellites by $z=2$, but significant for galaxies that stay centrals, with the final mass deviating on average by $50\%$. This is because whether a galaxy is star forming or not at high redshift correlates with the environment, where we showed that quenched galaxies tend to lie in underdense regions. 
We conclude that the environment of a galaxy at high redshift is not only important for it being quenched as shown by K23, but that it also plays a crucial role for the future development of the quenched galaxies at low redshifts. Thus, we suggest that relics of galaxies quenched at high redshifts can best be found in low mass group environments or isolated fields, albeit it might be possible for a high redshift quenched galaxy to be accreted onto a cluster and still remain as a compact quenched galaxy. As the quenched galaxies at high redshifts are very compact, relics of such galaxies should also still be significantly more compact than other quenched galaxies of comparable mass at present day. Nevertheless, as some quenched galaxies also merge with other, still star forming galaxies prior to $z=2$, it is also possible to find contributions of the quenched galaxies distributed in other galaxies, however, detecting those might be extremely difficult.

\section*{Acknowledgements}
This work was supported by the Deutsche Forschungsgemeinschaft (DFG, German Research Foundation) under Germany's Excellence Strategy - EXC-2094 - 390783311.
LCK acknowledges support by the DFG project nr. 516355818. 
The {\it Magneticum} simulations were performed at the Leibniz-Rechenzentrum with CPU time assigned to the Project {\it pr83li}.  We are especially grateful for the support by M. Petkova through the Computational Center for Particle and Astrophysics (C2PAP).

\bibliography{biblio}

\begin{thebibliography}{58}
\expandafter\ifx\csname natexlab\endcsname\relax\def\natexlab#1{#1}\fi

\bibitem[{{Adams} {et~al.}(2023){Adams}, {Conselice}, {Ferreira}, {Austin},
  {Trussler}, {Juod{\v{z}}balis}, {Wilkins}, {Caruana}, {Dayal}, {Verma}, \&
  {Vijayan}}]{adams:2023}
{Adams}, N.~J., {Conselice}, C.~J., {Ferreira}, L., {et~al.} 2023, \mnras, 518,
  4755

\bibitem[{{Arrabal Haro} {et~al.}(2023){Arrabal Haro}, {Dickinson},
  {Finkelstein}, {Kartaltepe}, {Donnan}, {Burgarella}, {Carnall}, {Cullen},
  {Dunlop}, {Fern{\'a}ndez}, {Fujimoto}, {Jung}, {Krips}, {Larson}, {Papovich},
  {P{\'e}rez-Gonz{\'a}lez}, {Amor{\'\i}n}, {Bagley}, {Buat}, {Casey},
  {Chworowsky}, {Cohen}, {Ferguson}, {Giavalisco}, {Huertas-Company},
  {Hutchison}, {Kocevski}, {Koekemoer}, {Lucas}, {McLeod}, {McLure}, {Pirzkal},
  {Seill{\'e}}, {Trump}, {Weiner}, {Wilkins}, \& {Zavala}}]{arrabal:2023}
{Arrabal Haro}, P., {Dickinson}, M., {Finkelstein}, S.~L., {et~al.} 2023, arXiv
  e-prints, arXiv:2303.15431

\bibitem[{{Beck} {et~al.}(2016){Beck}, {Murante}, {Arth}, {Remus}, {Teklu},
  {Donnert}, {Planelles}, {Beck}, {F{\"o}rster}, {Imgrund}, {Dolag}, \&
  {Borgani}}]{beck:2015}
{Beck}, A.~M., {Murante}, G., {Arth}, A., {et~al.} 2016, \mnras, 455, 2110

\bibitem[{{Bell} {et~al.}(2006){Bell}, {Naab}, {McIntosh}, {Somerville},
  {Caldwell}, {Barden}, {Wolf}, {Rix}, {Beckwith}, {Borch}, {H{\"a}ussler},
  {Heymans}, {Jahnke}, {Jogee}, {Koposov}, {Meisenheimer}, {Peng}, {Sanchez},
  \& {Wisotzki}}]{bell:2006}
{Bell}, E.~F., {Naab}, T., {McIntosh}, D.~H., {et~al.} 2006, \apj, 640, 241

\bibitem[{{Bezanson} {et~al.}(2009){Bezanson}, {van Dokkum}, {Tal},
  {Marchesini}, {Kriek}, {Franx}, \& {Coppi}}]{bezanson:2009}
{Bezanson}, R., {van Dokkum}, P.~G., {Tal}, T., {et~al.} 2009, \apj, 697, 1290

\bibitem[{{Bournaud} {et~al.}(2007){Bournaud}, {Jog}, \&
  {Combes}}]{bournaud:2007}
{Bournaud}, F., {Jog}, C.~J., \& {Combes}, F. 2007, \aap, 476, 1179

\bibitem[{{Carnall} {et~al.}(2023){Carnall}, {McLeod}, {McLure}, {Dunlop},
  {Begley}, {Cullen}, {Donnan}, {Hamadouche}, {Jewell}, {Jones}, {Pollock}, \&
  {Wild}}]{carnall:2023}
{Carnall}, A.~C., {McLeod}, D.~J., {McLure}, R.~J., {et~al.} 2023, \mnras, 520,
  3974

\bibitem[{{Chauke} {et~al.}(2019){Chauke}, {van der Wel}, {Pacifici},
  {Bezanson}, {Wu}, {Gallazzi}, {Straatman}, {Franx}, {Bari{\v{s}}i{\'c}},
  {Bell}, {van Houdt}, {Maseda}, {Muzzin}, {Sobral}, \&
  {Spilker}}]{chauke:2019}
{Chauke}, P., {van der Wel}, A., {Pacifici}, C., {et~al.} 2019, \apj, 877, 48

\bibitem[{{Dolag} {et~al.}(2009){Dolag}, {Borgani}, {Murante}, \&
  {Springel}}]{dolag:2009}
{Dolag}, K., {Borgani}, S., {Murante}, G., \& {Springel}, V. 2009, \mnras, 399,
  497

\bibitem[{{Dolag} {et~al.}(2004){Dolag}, {Jubelgas}, {Springel}, {Borgani}, \&
  {Rasia}}]{dolag:2004}
{Dolag}, K., {Jubelgas}, M., {Springel}, V., {Borgani}, S., \& {Rasia}, E.
  2004, \apjl, 606, L97

\bibitem[{{Dolag} {et~al.}(2017){Dolag}, {Mevius}, \& {Remus}}]{dolag:2017}
{Dolag}, K., {Mevius}, E., \& {Remus}, R.-S. 2017, Galaxies, 5, 35

\bibitem[{{Dolag} {et~al.}(2005){Dolag}, {Vazza}, {Brunetti}, \&
  {Tormen}}]{dolag:2005}
{Dolag}, K., {Vazza}, F., {Brunetti}, G., \& {Tormen}, G. 2005, \mnras, 364,
  753

\bibitem[{{Donnert} {et~al.}(2013){Donnert}, {Dolag}, {Brunetti}, \&
  {Cassano}}]{donnert:2013}
{Donnert}, J., {Dolag}, K., {Brunetti}, G., \& {Cassano}, R. 2013, \mnras, 429,
  3564

\bibitem[{{Fabjan} {et~al.}(2010){Fabjan}, {Borgani}, {Tornatore}, {Saro},
  {Murante}, \& {Dolag}}]{fabjan:2010}
{Fabjan}, D., {Borgani}, S., {Tornatore}, L., {et~al.} 2010, \mnras, 401, 1670

\bibitem[{{Franx} {et~al.}(2008){Franx}, {van Dokkum}, {F{\"o}rster Schreiber},
  {Wuyts}, {Labb{\'e}}, \& {Toft}}]{franx:2008}
{Franx}, M., {van Dokkum}, P.~G., {F{\"o}rster Schreiber}, N.~M., {et~al.}
  2008, \apj, 688, 770

\bibitem[{{Fudamoto} {et~al.}(2022){Fudamoto}, {Inoue}, \&
  {Sugahara}}]{fudamoto:2022}
{Fudamoto}, Y., {Inoue}, A.~K., \& {Sugahara}, Y. 2022, \apjl, 938, L24

\bibitem[{{Fudamoto} {et~al.}(2021){Fudamoto}, {Oesch}, {Schouws}, {Stefanon},
  {Smit}, {Bouwens}, {Bowler}, {Endsley}, {Gonzalez}, {Inami}, {Labbe},
  {Stark}, {Aravena}, {Barrufet}, {da Cunha}, {Dayal}, {Ferrara}, {Graziani},
  {Hodge}, {Hutter}, {Li}, {De Looze}, {Nanayakkara}, {Pallottini}, {Riechers},
  {Schneider}, {Ucci}, {van der Werf}, \& {White}}]{fudamoto:2021}
{Fudamoto}, Y., {Oesch}, P.~A., {Schouws}, S., {et~al.} 2021, \nat, 597, 489

\bibitem[{{Glazebrook} {et~al.}(2023){Glazebrook}, {Nanayakkara}, {Schreiber},
  {Lagos}, {Kawinwanichakij}, {Jacobs}, {Chittenden}, {Brammer}, {Kacprzak},
  {Labbe}, {Marchesini}, {Marsan}, {Oesch}, {Papovich}, {Remus}, {Tran},
  {Esdaile}, \& {Chandro Gomez}}]{glazebrook:2023}
{Glazebrook}, K., {Nanayakkara}, T., {Schreiber}, C., {et~al.} 2023, arXiv
  e-prints, arXiv:2308.05606

\bibitem[{{Haardt} \& {Madau}(2001)}]{haardt:2001}
{Haardt}, F. \& {Madau}, P. 2001, in Clusters of Galaxies and the High Redshift
  Universe Observed in X-rays, ed. D.~M. {Neumann} \& J.~T.~V. {Tran}, 64

\bibitem[{{Harikane} {et~al.}(2023{\natexlab{a}}){Harikane}, {Nakajima},
  {Ouchi}, {Umeda}, {Isobe}, {Ono}, {Xu}, \& {Zhang}}]{harikane:2023spec}
{Harikane}, Y., {Nakajima}, K., {Ouchi}, M., {et~al.} 2023{\natexlab{a}}, arXiv
  e-prints, arXiv:2304.06658

\bibitem[{{Harikane} {et~al.}(2023{\natexlab{b}}){Harikane}, {Ouchi}, {Oguri},
  {Ono}, {Nakajima}, {Isobe}, {Umeda}, {Mawatari}, \& {Zhang}}]{harikane:2023}
{Harikane}, Y., {Ouchi}, M., {Oguri}, M., {et~al.} 2023{\natexlab{b}}, \apjs,
  265, 5

\bibitem[{{Hartley} {et~al.}(2023){Hartley}, {Nelson}, {Suess}, {Garcia},
  {Park}, {Hernquist}, {Bezanson}, {Nevin}, {Pillepich}, {Schechter},
  {Terrazas}, {Torrey}, {Wellons}, {Whitaker}, \& {Williams}}]{hartley:2023}
{Hartley}, A.~I., {Nelson}, E.~J., {Suess}, K.~A., {et~al.} 2023, \mnras, 522,
  3138

\bibitem[{{Hilz} {et~al.}(2012){Hilz}, {Naab}, {Ostriker}, {Thomas}, {Burkert},
  \& {Jesseit}}]{hilz:2012}
{Hilz}, M., {Naab}, T., {Ostriker}, J.~P., {et~al.} 2012, \mnras, 425, 3119

\bibitem[{{Ito} {et~al.}(2023){Ito}, {Valentino}, {Brammer}, {Faisst},
  {Gillman}, {Gomez-Guijarro}, {Gould}, {Heintz}, {Ilbert}, {Kragh Jespersen},
  {Kokorev}, {Kubo}, {Magdis}, {McPartland}, {Onodera}, {Rizzo}, {Tanaka},
  {Toft}, {Vijayan}, {Weaver}, {Whitaker}, \& {Wright}}]{ito:2023}
{Ito}, K., {Valentino}, F., {Brammer}, G., {et~al.} 2023, arXiv e-prints,
  arXiv:2307.06994

\bibitem[{{Kakimoto} {et~al.}(2023){Kakimoto}, {Tanaka}, {Onodera},
  {Shimakawa}, {Wu}, {Gould}, {Ito}, {Jin}, {Kubo}, {Suzuki}, {Toft},
  {Valentino}, \& {Yabe}}]{kakimoto:2023}
{Kakimoto}, T., {Tanaka}, M., {Onodera}, M., {et~al.} 2023, arXiv e-prints,
  arXiv:2308.15011

\bibitem[{{Karademir} {et~al.}(2019){Karademir}, {Remus}, {Burkert}, {Dolag},
  {Hoffmann}, {Moster}, {Steinwandel}, \& {Zhang}}]{karademir:2019}
{Karademir}, G.~S., {Remus}, R.-S., {Burkert}, A., {et~al.} 2019, \mnras, 487,
  318

\bibitem[{{Komatsu} {et~al.}(2011){Komatsu}, {Smith}, {Dunkley}, {Bennett},
  {Gold}, {Hinshaw}, {Jarosik}, {Larson}, {Nolta}, \& {Page}}]{komatsu:2011}
{Komatsu}, E., {Smith}, K.~M., {Dunkley}, J., {et~al.} 2011, \apjs, 192, 18

\bibitem[{{Kurinchi-Vendhan} {et~al.}(2023){Kurinchi-Vendhan}, {Farcy},
  {Hirschmann}, \& {Valentino}}]{kurinchi:2023}
{Kurinchi-Vendhan}, S., {Farcy}, M., {Hirschmann}, M., \& {Valentino}, F. 2023,
  arXiv e-prints, arXiv:2310.03083

\bibitem[{{Lange} {et~al.}(2015){Lange}, {Driver}, {Robotham}, {Kelvin},
  {Graham}, {Alpaslan}, {Andrews}, {Baldry}, {Bamford}, {Bland-Hawthorn},
  {Brough}, {Cluver}, {Conselice}, {Davies}, {Haeussler}, {Konstantopoulos},
  {Loveday}, {Moffett}, {Norberg}, {Phillipps}, {Taylor},
  {L{\'o}pez-S{\'a}nchez}, \& {Wilkins}}]{lange:2015}
{Lange}, R., {Driver}, S.~P., {Robotham}, A. S.~G., {et~al.} 2015, \mnras, 447,
  2603

\bibitem[{{Lelli} {et~al.}(2023){Lelli}, {Zhang}, {Bisbas}, {Lin},
  {Papadopoulos}, {Schombert}, {Di Teodoro}, {Marasco}, \&
  {McGaugh}}]{lelli:2023}
{Lelli}, F., {Zhang}, Z.-Y., {Bisbas}, T.~G., {et~al.} 2023, \aap, 672, A106

\bibitem[{{Long} {et~al.}(2023){Long}, {Antwi-Danso}, {Lambrides}, {Lovell},
  {de la Vega}, {Valentino}, {Zavala}, {Casey}, {Wilkins}, {Yung}, {Arrabal
  Haro}, {Bagley}, {Bisigello}, {Chworowsky}, {Cooper}, {Cooper}, {Cooray},
  {Croton}, {Dickinson}, {Finkelstein}, {Franco}, {Gould}, {Hirschmann},
  {Hutchison}, {Kartaltepe}, {Kocevski}, {Koekemoer}, {Lucas}, {McKinney},
  {Papovich}, {Perez-Gonzalez}, {Pirzkal}, \& {Santini}}]{long:2023}
{Long}, A.~S., {Antwi-Danso}, J., {Lambrides}, E.~L., {et~al.} 2023, arXiv
  e-prints, arXiv:2305.04662

\bibitem[{{Mart{\'\i}n-Navarro} {et~al.}(2022){Mart{\'\i}n-Navarro}, {Shankar},
  \& {Mezcua}}]{martinnavarro:2022}
{Mart{\'\i}n-Navarro}, I., {Shankar}, F., \& {Mezcua}, M. 2022, \mnras, 513,
  L10

\bibitem[{{Naab} {et~al.}(2009){Naab}, {Johansson}, \& {Ostriker}}]{naab:2009}
{Naab}, T., {Johansson}, P.~H., \& {Ostriker}, J.~P. 2009, \apjl, 699, L178

\bibitem[{{Nanayakkara} {et~al.}(2022){Nanayakkara}, {Glazebrook}, {Jacobs},
  {Schreiber}, {Brammer}, {Esdaile}, {Kacprzak}, {Labbe}, {Lagos},
  {Marchesini}, {Marsan}, {Nateghi}, {Oesch}, {Papovich}, {Remus}, \&
  {Tran}}]{nanayakkara:2022}
{Nanayakkara}, T., {Glazebrook}, K., {Jacobs}, C., {et~al.} 2022, arXiv
  e-prints, arXiv:2212.11638

\bibitem[{{Nelson} {et~al.}(2023){Nelson}, {Suess}, {Bezanson}, {Price}, {van
  Dokkum}, {Leja}, {Wang}, {Whitaker}, {Labb{\'e}}, {Barrufet}, {Brammer},
  {Eisenstein}, {Gibson}, {Hartley}, {Johnson}, {Heintz}, {Mathews}, {Miller},
  {Oesch}, {Sandles}, {Setton}, {Speagle}, {Tacchella}, {Tadaki}, {{\"U}bler},
  \& {Weaver}}]{nelson:2023}
{Nelson}, E.~J., {Suess}, K.~A., {Bezanson}, R., {et~al.} 2023, \apjl, 948, L18

\bibitem[{{Pearson} {et~al.}(2018){Pearson}, {Wang}, {Hurley}, {Ma{\l}ek},
  {Buat}, {Burgarella}, {Farrah}, {Oliver}, {Smith}, \& {van der
  Tak}}]{pearson:2018}
{Pearson}, W.~J., {Wang}, L., {Hurley}, P.~D., {et~al.} 2018, \aap, 615, A146

\bibitem[{{Remus} {et~al.}(2023){Remus}, {Dolag}, \&
  {Dannerbauer}}]{remus:2023}
{Remus}, R.-S., {Dolag}, K., \& {Dannerbauer}, H. 2023, \apj, 950, 191

\bibitem[{{Remus} {et~al.}(2017){Remus}, {Dolag}, {Naab}, {Burkert},
  {Hirschmann}, {Hoffmann}, \& {Johansson}}]{remus:2017}
{Remus}, R.-S., {Dolag}, K., {Naab}, T., {et~al.} 2017, \mnras, 464, 3742

\bibitem[{{Rizzo} {et~al.}(2023){Rizzo}, {Roman-Oliveira}, {Fraternali},
  {Frickmann}, {Valentino}, {Brammer}, {Zanella}, {Kokorev}, {Popping},
  {Whitaker}, {Kohandel}, {Magdis}, {Di Mascolo}, {Ikeda}, {Jin}, \&
  {Toft}}]{rizzo:2023}
{Rizzo}, F., {Roman-Oliveira}, F., {Fraternali}, F., {et~al.} 2023, arXiv
  e-prints, arXiv:2303.16227

\bibitem[{{Roman-Oliveira} {et~al.}(2023){Roman-Oliveira}, {Fraternali}, \&
  {Rizzo}}]{roman:2023}
{Roman-Oliveira}, F., {Fraternali}, F., \& {Rizzo}, F. 2023, \mnras, 521, 1045

\bibitem[{{Salmon} {et~al.}(2015){Salmon}, {Papovich}, {Finkelstein}, {Tilvi},
  {Finlator}, {Behroozi}, {Dahlen}, {Dav{\'e}}, {Dekel}, {Dickinson},
  {Ferguson}, {Giavalisco}, {Long}, {Lu}, {Mobasher}, {Reddy}, {Somerville}, \&
  {Wechsler}}]{salmon:2015}
{Salmon}, B., {Papovich}, C., {Finkelstein}, S.~L., {et~al.} 2015, \apj, 799,
  183

\bibitem[{{Schreiber} {et~al.}(2018){Schreiber}, {Glazebrook}, {Nanayakkara},
  {Kacprzak}, {Labb{\'e}}, {Oesch}, {Yuan}, {Tran}, {Papovich}, {Spitler}, \&
  {Straatman}}]{schreiber:2018}
{Schreiber}, C., {Glazebrook}, K., {Nanayakkara}, T., {et~al.} 2018, \aap, 618,
  A85

\bibitem[{{Schulze} {et~al.}(2018){Schulze}, {Remus}, {Dolag}, {Burkert},
  {Emsellem}, \& {van de Ven}}]{schulze:2018}
{Schulze}, F., {Remus}, R.-S., {Dolag}, K., {et~al.} 2018, \mnras, 480, 4636

\bibitem[{{Springel}(2005)}]{springel:2005}
{Springel}, V. 2005, \mnras, 364, 1105

\bibitem[{{Springel} \& {Hernquist}(2003)}]{springel:2003}
{Springel}, V. \& {Hernquist}, L. 2003, \mnras, 339, 289

\bibitem[{{Springel} {et~al.}(2005){Springel}, {White}, {Jenkins}, {Frenk},
  {Yoshida}, {Gao}, {Navarro}, {Thacker}, {Croton}, {Helly}, {Peacock}, {Cole},
  {Thomas}, {Couchman}, {Evrard}, {Colberg}, \& {Pearce}}]{springel:2005b}
{Springel}, V., {White}, S. D.~M., {Jenkins}, A., {et~al.} 2005, \nat, 435, 629

\bibitem[{{Springel} {et~al.}(2001){Springel}, {White}, {Tormen}, \&
  {Kauffmann}}]{springel:2001}
{Springel}, V., {White}, S.~D.~M., {Tormen}, G., \& {Kauffmann}, G. 2001,
  \mnras, 328, 726

\bibitem[{{Steinborn} {et~al.}(2015){Steinborn}, {Dolag}, {Hirschmann},
  {Prieto}, \& {Remus}}]{steinborn:2015}
{Steinborn}, L.~K., {Dolag}, K., {Hirschmann}, M., {Prieto}, M.~A., \& {Remus},
  R.-S. 2015, \mnras, 448, 1504

\bibitem[{{Tanaka} {et~al.}(2023){Tanaka}, {Shimasaku}, {Tacchella}, {Ando},
  {Ito}, {Yesuf}, \& {Matsui}}]{tanaka:2023}
{Tanaka}, T.~S., {Shimasaku}, K., {Tacchella}, S., {et~al.} 2023, arXiv
  e-prints, arXiv:2307.14235

\bibitem[{{Teklu} {et~al.}(2015){Teklu}, {Remus}, {Dolag}, {Beck}, {Burkert},
  {Schmidt}, {Schulze}, \& {Steinborn}}]{teklu:2015}
{Teklu}, A.~F., {Remus}, R.-S., {Dolag}, K., {et~al.} 2015, \apj, 812, 29

\bibitem[{{Tornatore} {et~al.}(2007){Tornatore}, {Borgani}, {Dolag}, \&
  {Matteucci}}]{tornatore:2007}
{Tornatore}, L., {Borgani}, S., {Dolag}, K., \& {Matteucci}, F. 2007, \mnras,
  382, 1050

\bibitem[{{Tornatore} {et~al.}(2004){Tornatore}, {Borgani}, {Matteucci},
  {Recchi}, \& {Tozzi}}]{tornatore:2004}
{Tornatore}, L., {Borgani}, S., {Matteucci}, F., {Recchi}, S., \& {Tozzi}, P.
  2004, \mnras, 349, L19

\bibitem[{{Tsukui} \& {Iguchi}(2021)}]{tsukui:2021}
{Tsukui}, T. \& {Iguchi}, S. 2021, Science, 372, 1201

\bibitem[{{Tsukui} {et~al.}(2023){Tsukui}, {Wisnioski}, {Krumholz}, \&
  {Battisti}}]{tsukui:2023}
{Tsukui}, T., {Wisnioski}, E., {Krumholz}, M.~R., \& {Battisti}, A. 2023,
  \mnras, 523, 4654

\bibitem[{{van der Wel} {et~al.}(2014){van der Wel}, {Franx}, {van Dokkum},
  {Skelton}, {Momcheva}, {Whitaker}, {Brammer}, {Bell}, {Rix}, {Wuyts},
  {Ferguson}, {Holden}, {Barro}, {Koekemoer}, {Chang}, {McGrath},
  {H{\"a}ussler}, {Dekel}, {Behroozi}, {Fumagalli}, {Leja}, {Lundgren},
  {Maseda}, {Nelson}, {Wake}, {Patel}, {Labb{\'e}}, {Faber}, {Grogin}, \&
  {Kocevski}}]{vdwel:2014}
{van der Wel}, A., {Franx}, M., {van Dokkum}, P.~G., {et~al.} 2014, \apj, 788,
  28

\bibitem[{{Wiersma} {et~al.}(2009){Wiersma}, {Schaye}, \&
  {Smith}}]{wiersma:2009}
{Wiersma}, R.~P.~C., {Schaye}, J., \& {Smith}, B.~D. 2009, \mnras, 393, 99

\bibitem[{{Yi} {et~al.}(2005){Yi}, {Yoon}, {Kaviraj}, {Deharveng}, {Rich},
  {Salim}, {Boselli}, {Lee}, {Ree}, {Sohn}, {Rey}, {Lee}, {Rhee}, {Bianchi},
  {Byun}, {Donas}, {Friedman}, {Heckman}, {Jelinsky}, {Madore}, {Malina},
  {Martin}, {Milliard}, {Morrissey}, {Neff}, {Schiminovich}, {Siegmund},
  {Small}, {Szalay}, {Jee}, {Kim}, {Barlow}, {Forster}, {Welsh}, \&
  {Wyder}}]{yi:2005}
{Yi}, S.~K., {Yoon}, S.~J., {Kaviraj}, S., {et~al.} 2005, \apjl, 619, L111

\bibitem[{{Zhang} {et~al.}(2023){Zhang}, {Li}, {Leja}, {Whitaker}, {Nersesian},
  {Bezanson}, \& {van der Wel}}]{zhang:2023}
{Zhang}, J., {Li}, Y., {Leja}, J., {et~al.} 2023, \apj, 952, 6

\end{thebibliography}
\bibliographystyle{aa}

\label{lastpage}
\end{document}